\documentclass[journal]{IEEEtran}
\usepackage{cite}
\usepackage{amsmath,amssymb,amsfonts}
\usepackage{algorithmic}
\usepackage{graphicx}
\usepackage{textcomp}
\usepackage{xcolor}
\usepackage{epsfig}
\usepackage{multirow}
\usepackage{dblfloatfix}
\usepackage{subfigure}
\usepackage{caption}
\def\BibTeX{{\rm B\kern-.05em{\sc i\kern-.025em b}\kern-.08em
    T\kern-.1667em\lower.7ex\hbox{E}\kern-.125emX}}
\begin{document}

\title{VWA: Hardware Efficient Vectorwise Accelerator for Convolutional Neural Network}

\author{Kuo-Wei~Chang,~
	Tian-Sheuan~Chang,~\IEEEmembership{Senior Member,~IEEE}%
\thanks{
© 2020 IEEE.  Personal use of this material is permitted.  Permission from IEEE must be obtained for all other uses, in any current or future media, including reprinting/republishing this material for advertising or promotional purposes, creating new collective works, for resale or redistribution to servers or lists, or reuse of any copyrighted component of this work in other works.\\
K. Chang and T. Chang, "VWA: Hardware Efficient Vectorwise Accelerator for Convolutional Neural Network," in IEEE Transactions on Circuits and Systems I: Regular Papers, vol. 67, no. 1, pp. 145-154, Jan. 2020, doi: 10.1109/TCSI.2019.2942529.
}
}

\maketitle

\begin{abstract}
Hardware accelerators for convolution neural networks (CNNs) enable real-time applications of artificial intelligence technology. However, most of the existing designs suffer from low hardware utilization or high area cost due to complex dataflow. This paper proposes a hardware efficient vectorwise CNN accelerator that adopts a 3$\times$3 filter optimized systolic array using 1-D broadcast dataflow to generate partial sum. This enables easy reconfiguration for different kinds of kernels with interleaved input or elementwise input dataflow. This simple and regular data flow results in low area cost while attains high hardware utilization. The presented design achieves 99\%, 97\%, 93.7\%, 94\% hardware utilization for VGG-16, ResNet-34, GoogLeNet, and Mobilenet, respectively. Hardware implementation with TSMC 40nm technology  takes 266.9K NAND gate count and 191KB SRAM to support 168GOPS throughput and consumes only 154.98mW when running at 500MHz operating frequency, which has superior area and power efficiency than other designs. 

\end{abstract}


\section{Introduction}
Deep convolution neural networks (DCNNs) have been widely used in computer vision tasks, such as recognition\cite{1,2,3,4,5}, detection\cite{6,7,8,9,10}, and autonomous vehicles during recent years for its significant improvement over traditional approaches. However, the computation of \textcolor{black}{CNNs} demands a lot of multiplications and accumulations (MACs) and millions of data amount per layer that prohibits its wide usage in real-time applications. Thus, hardware acceleration for DCNNs is demanded to satisfy the computation and bandwidth requirements \textcolor{black}{under} the real time constraint. 

Various hardware accelerators have been proposed recently [11-23, 33-38].\cite{Diannao1,Diannao2,Diannao3,Diannao4,Diannao5} minimize the communication traffic by a tiling strategy and internal buffer. However, the modern DCNNs are too large to store all weights into the on-chip memory. \cite{eyeriss}  adopts a spatial array architecture and row stationary data flow that helps maximize data reuse but leads to high processing element (PE) cost due to large local storage. In addition, its fixed PE configuration results in low hardware utilization. Its improved version \cite{eyerissv2} has higher hardware utilization and supports sparse CNNs by clusters to increase throughput, but \textcolor{black}{needs} large area overhead. \cite{dna} proposes a systolic array architecture with full reconfigurations for different convolutional kernels. It achieves high hardware utilization but needs significant PE area cost due to complex control overhead. The design in \cite{envision} is precision scalable to achieve low power \textcolor{black}{consumption under different application scenarios.} \cite{yj} uses a filter-type structure that can be reconfigublack to fit different needs but this filter-like data flow limits the flexibility of reconfigurations. The streaming architecture in \cite{mcc} blackuces data movement through filter reuse and input reuse to optimize energy efficiency, but it has low hardware utilization beyond optimized 3$\times$3 architectures. The systolic array architecture in \cite{3dtile}  is also a reconfigurable design to fit different convolution kernels but with a propagated input data flow that results in long latency and low hardware utilization. \cite{DTCNN} proposes an accelerator for image segmentation which supports dilated and transposed convolution. In addition, it cuts down the blackundant zero computations for higher throughput. \textcolor{black}{\cite{MAERI} provides near non-blocking communication via reconfigurable links with high bandwidth to support efficient irregular dataflow mapping and provide high utilization of computing units. \cite{R1} proposes a low precision design that uses heterogeneous representation and significant bits securing encoding which can exploit the advantage of both efficiency of fixed-point and flexibility of floating-point to optimize their PE architecture and consume less power and area. \cite{R2} adopts different combinations of input, weight and output data flow to optimize their design. The data flow in \cite{R3} inspiblack from the fully-connected layer can accelerate convolutional processes and parametrize design from low memory accesses to low latency systems. \cite{R4} is a computing in memory design that adopts SRAM array for vector matrix multiplication for low power deep learning applications. \cite{R5} implements convolutions of different kernel sizes with multiple parallel fast FIR based convolution units.} Although existing designs can support most of kernel sizes and stride lengths, they suffer from low hardware utilization or high hardware cost due to constrained data flow or complex PE structures \textcolor{black}{for such reconfiguration requirements.}
 \begin{figure}[t]
	\centering{\includegraphics[height=28mm]{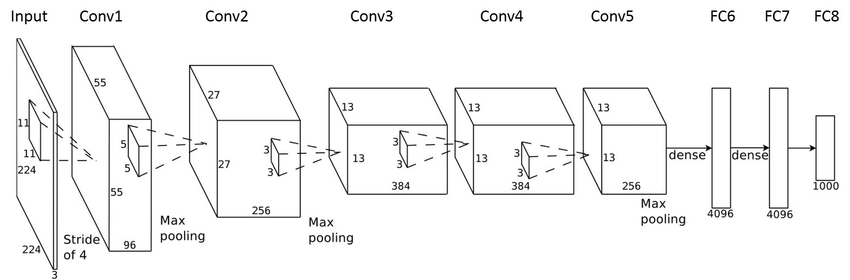}}
	\caption{A typical CNN model, Alexnet\cite{alexnet}.}
	\label{fig:alexnet}
\end{figure}

To achieve high hardware utilization with low area overhead, this paper proposes a hardware efficient vectorwise accelerator \textcolor{black}{based on a PE array} for CNN. This design \textcolor{black}{adopts} broadcasted vectorwise input and weight \textcolor{black}{for the PE array} to avoid local weight storage. The array is partitioned into blocks and optimized for the widely used 3$\times$3 convolutions but also easily reconfigublack for other kernel size due to the broadcasted data flow. The PE array has high hardware utilization even for the non-unit stride case by the interleaved input. This design can also support 1$\times$1 convolution \textcolor{black}{commonly used} in the low complexity network with the elementwise input dataflow. \textcolor{black}{Even with above versatile reconfigurations, the proposed design still has high hardware efficiency compablack to other designs.}

%
%
%
%

The rest of the paper is organized as following. Section II gives a short overview of the CNN algorithm. Section III presents the proposed architecture. Section IV shows the details of the data flow for all kinds of convolutional filters supported in our architecture. The implementation results and comparisons are shown in Section V. Finally, we conclude this paper in section VI.

\begin{figure}[t]
	\centering{\includegraphics[height=55mm]{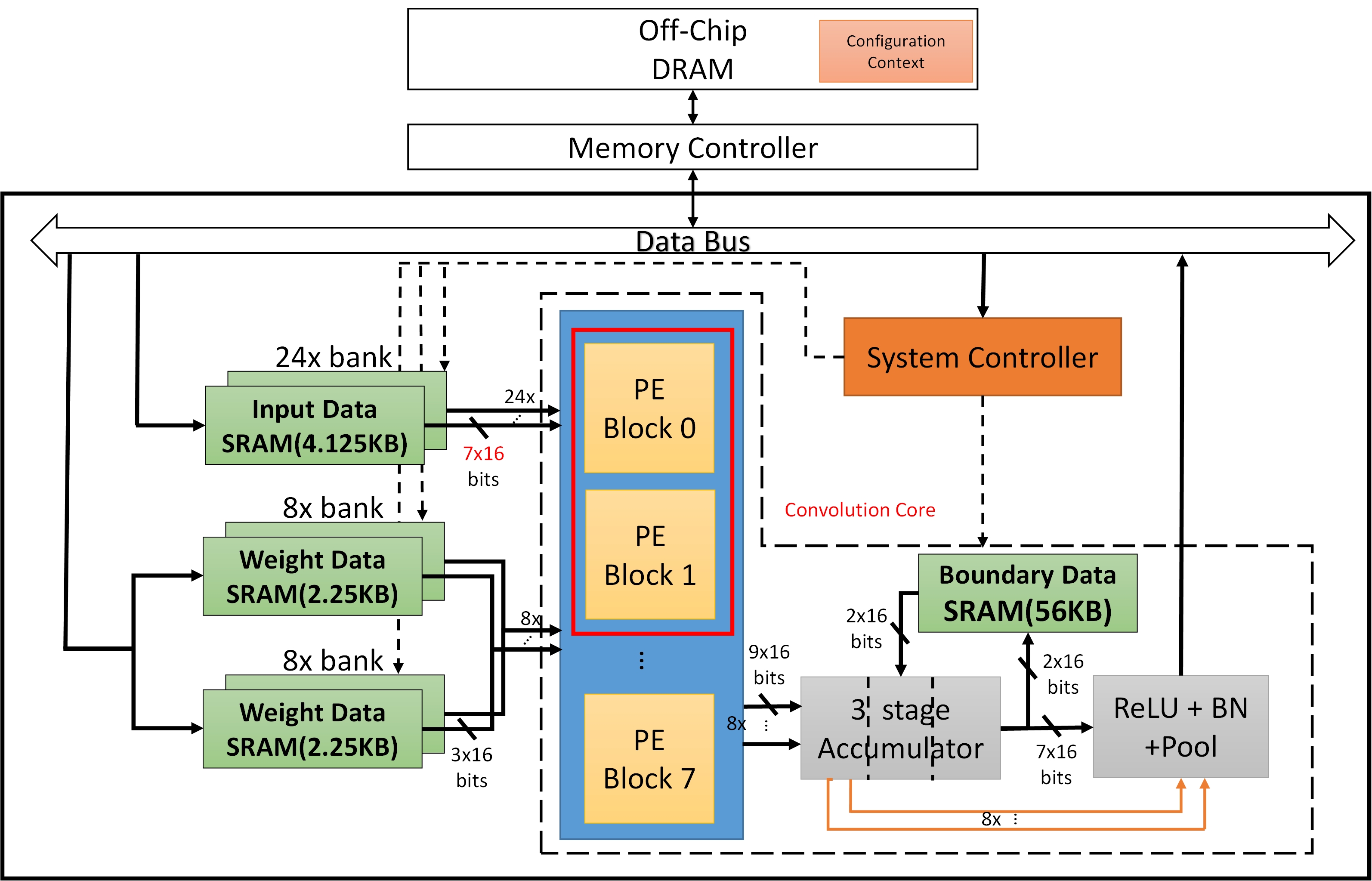}}
	\caption{The proposed system architecture.}
	\label{fig:Top}
\end{figure}

\section{Overview of CNN}
Fig.~\ref{fig:alexnet} shows a typical DCNN model, called AlexNet\cite{alexnet}, that classifies a 224$\times$224 RGB image into one of 1000 categories. A typical CNN structure consists of the convolutional layers,  pooling layers, fully-connected layers, and the activation function. The convolutional layers transform images into highly abstract representations called feature maps. The pooling layers execute down-sampling operations to blackuce feature map size and improve the translational invariance of features. In the state-of-art networks, the pooling layers could be replaced by layers of 3$\times$3 convolution with stride two. The commonly used activation function like rectified linear unit (ReLU) adds non-linearity after the convolutional layers or the fully-connected layers to improve feature representations. The fully-connected layers classify the image based on those extracted features, which is a pure neural network with a lot of weight numbers. 

The convolutional layer applies a filter kernel to the whole image, which can be defined as
\begin{equation}
\begin{split}
O(b,c,r,w)=Act(\sum_{u=0}^{U-1}\sum_{i=0}^{I-1}\sum_{j=0}^{J-1}{F(b,u,Sr+i,Sw+j)}\\
\times W(c,u,i,j)+B(u))
\end{split}
\end{equation}
\textcolor{black}{where $\boldsymbol{O}$, $\boldsymbol{F}$, $\boldsymbol{W}$, and $\boldsymbol{B}$ are the matrices of output feature maps, input feature maps, weight, and bias, respectively. In which, $Act$ is the activation function, such as ReLU, $b$ is the image batch size, $c$ is the number of filters, $r$ is the output row size, $w$ is the output column size, $u$ is the input channel, $i$ is the weight row size, $j$ is the weight column size, and $S$ is the stride length. For} this whole image operation, the computation needs $c \times u \times i \times j \times r \times w \times b$ MACs for one layer. The computation cost of the whole model is increased significantly  with modern deeper and wider models, which will be over tens or hundblacks of giga or tera operations per second for models like ResNet and GoogLeNet\textcolor{black}{\cite{39}}.

\begin{figure}[t]
	\centering{\includegraphics[height=90mm]{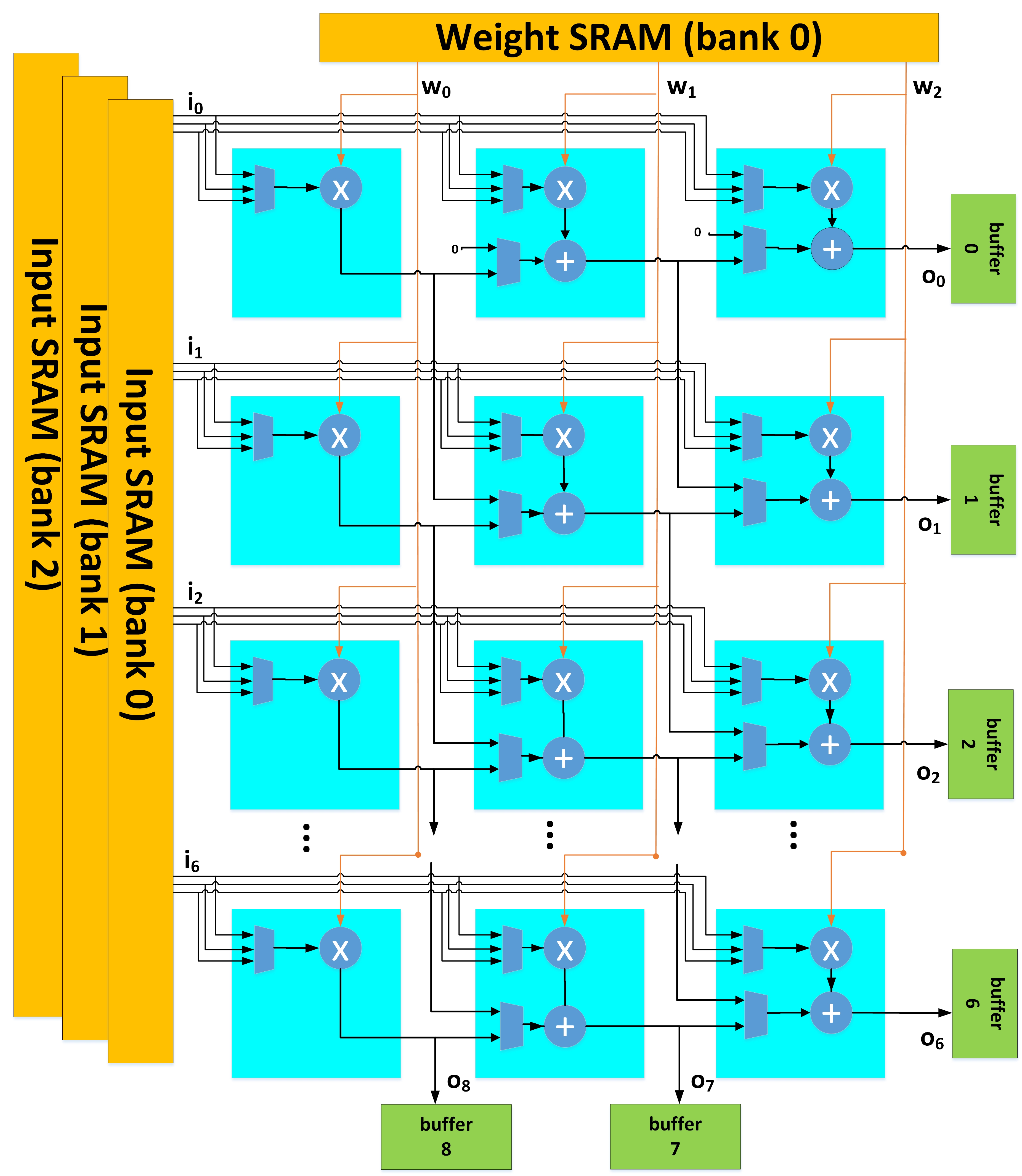}}
	\caption{\textcolor{black}{The architecture of the PE block. In which, 3$\times$7 input data and 3 weight data are broadcasted to a 7$\times$3 MAC array and generate 9 output data.}}
	\label{fig:PE}
\end{figure}
\section{System Architecture}

\subsection{Overview}
Fig.~\ref{fig:Top} shows the proposed system architecture, which consists of eight PE blocks, three-stage accumulators and a boundary data SRAM buffer for convolution computation, a post-processing module for ReLU, batch-normalization and pooling, and multi-bank input and weight SRAM buffers for continuous data access from external memory. The weight data buffer is a ping-pong buffer, which is divided into eight banks for eight PEs, respectively. The input buffer is divided into 24 banks for eight PEs. Each PE block will connect to three input SRAM banks to support different filter sizes.

The proposed design will first load a tile of input map and weight data to the local SRAM buffers for data reuse. Each PE block will use one channel of input map and one weight filter to compute partial convolution results. With eight PEs, we will be able to process eight input channels and eight filters at the same time. The partial convolution results from all PE blocks will be stoblack and accumulated locally through the three stage accumulators to avoid external memory access. During the above accumulation, partial results at the \textcolor{black}{top and bottom} boundary of tiles will be also stoblack in the boundary buffer to avoid repeated computation at the tile boundary. The final convolutional result will be processed by the post-processing module for ReLU, batch-normalization, and pooling operations, and then stoblack to external memory. 

\begin{figure}[t]
	\centering{\includegraphics[height=50mm]{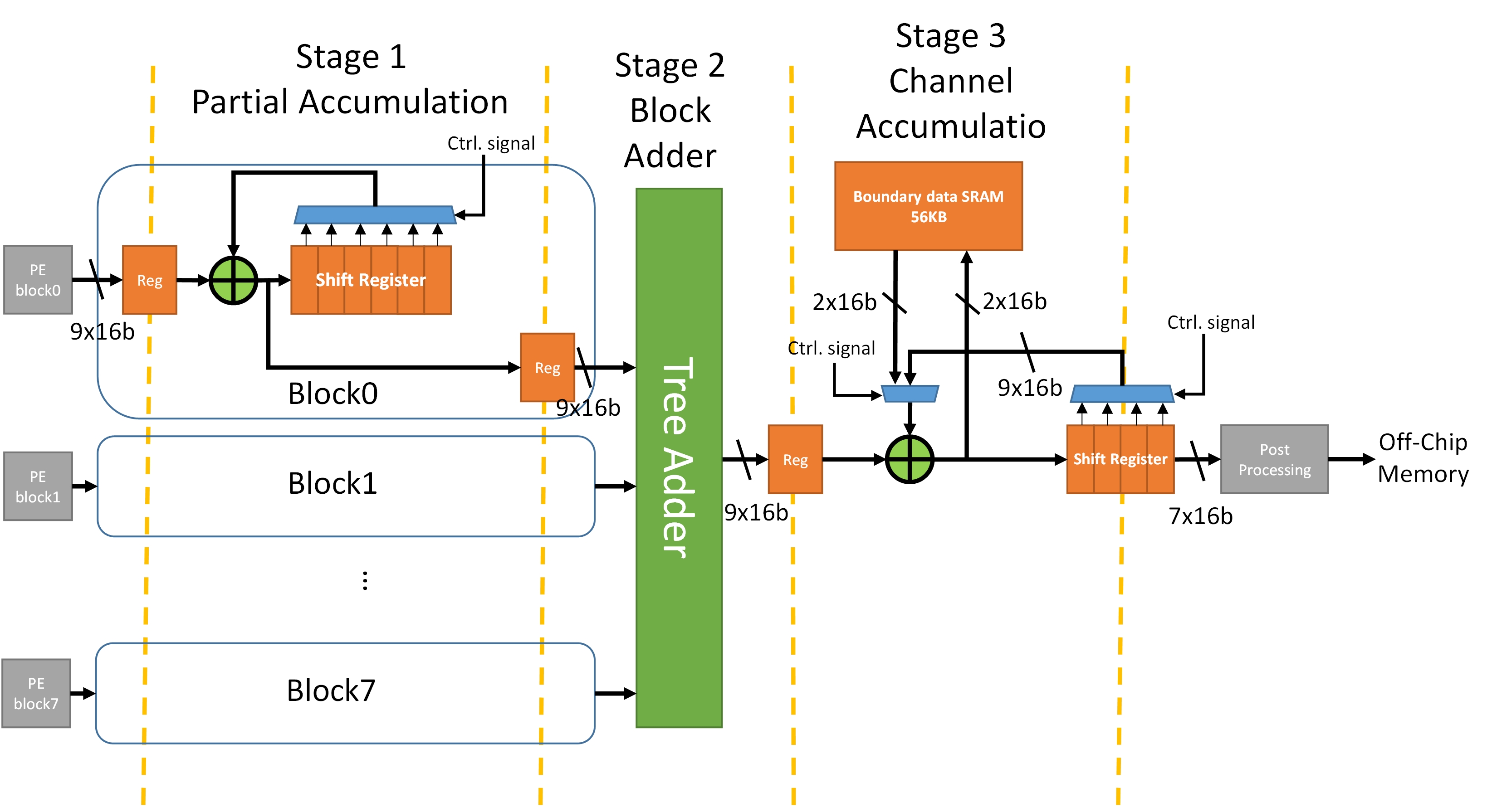}}
	\caption{The pipelined accumulator.}
	\label{fig:acc}
\end{figure}

\subsection{\textcolor{black}{PE Blocks}}
Fig.~\ref{fig:PE} illustrates the 7$\times$3 MAC array design of one PE block. The PE block has 3$\times$7 inputs broadcasted horizontally and three weights broadcasted vertically that are multiplied together and then summed along horizontal direction for an 1$\times$1 filter or diagonal direction for other filter sizes. \textcolor{black}{To support different convolution filters, each input row has three inputs so that each MAC can select the desiblack input via a 3-to-1 multiplexer.} This design chooses a three-column width \textcolor{black}{for weights} to optimize computations of 3$\times$3 convolutions and avoid long critical path. With this proposed data flow, we can maximize hardware utilization while keep data flow regular for lower area cost. 

Beyond the above block level reconfiguration, the proposed eight PE blocks can also be reconfigublack as two configurations, (8, 7, 3): 8 blocks, and 7 rows by 3 columns per block, or (4, 14, 3): 4 blocks, and 14 rows by 3 columns per block that combines two PE blocks into one as in Fig.~\ref{fig:Top}. This array level reconfiguration can support different input map sizes while maximize hardware utilization. For example, the PE array is reconfigublack as (4, 14, 3) in \textcolor{black}{the first layer of the models for image recognition, such as AlexNet, VGGNet, and ResNet} because the input map \textcolor{black}{only consists of} three channels with R, G, and B. \textcolor{black}{But the PE array can be} reconfigublack as (8, 7, 3) in \textcolor{black}{other layers due to wider channel numbers.}

\subsection{Accumulator}
Fig.~\ref{fig:acc} shows the accumulator that consists of a  three-stage pipeline structure to accumulate PE output along the channel dimension. For clarity of explanation, we will assume a 3$\times$3 filter in the rest of this subsection. \textcolor{black}{As shown in Fig.~\ref{fig:PE}}, each PE block has nine outputs, which are partial sums of one filter column for one channel. For a 3$\times$3 filter, three partial sums of the same channel will be accumulated at the first stage. Then, the filteblack output from eight channels (that is, from eight PE blocks) will be summed together with a tree adder at the second stage. For channel number larger than eight, they are divided into groups of eight channel processing in the PE blocks. Then, these different groups of channels will be accumulated at the third stage to generate the convolution output. If these output data are at the tile boundary, they will be stoblack in the global boundary data SRAM and then accumulated later with the output from the neighboring tile output. This accumulator is reconfigurable to fit the needs of different convolution kernels as illustrated in the later section.
\begin{figure}[t]
	\centering{\includegraphics[height=40mm]{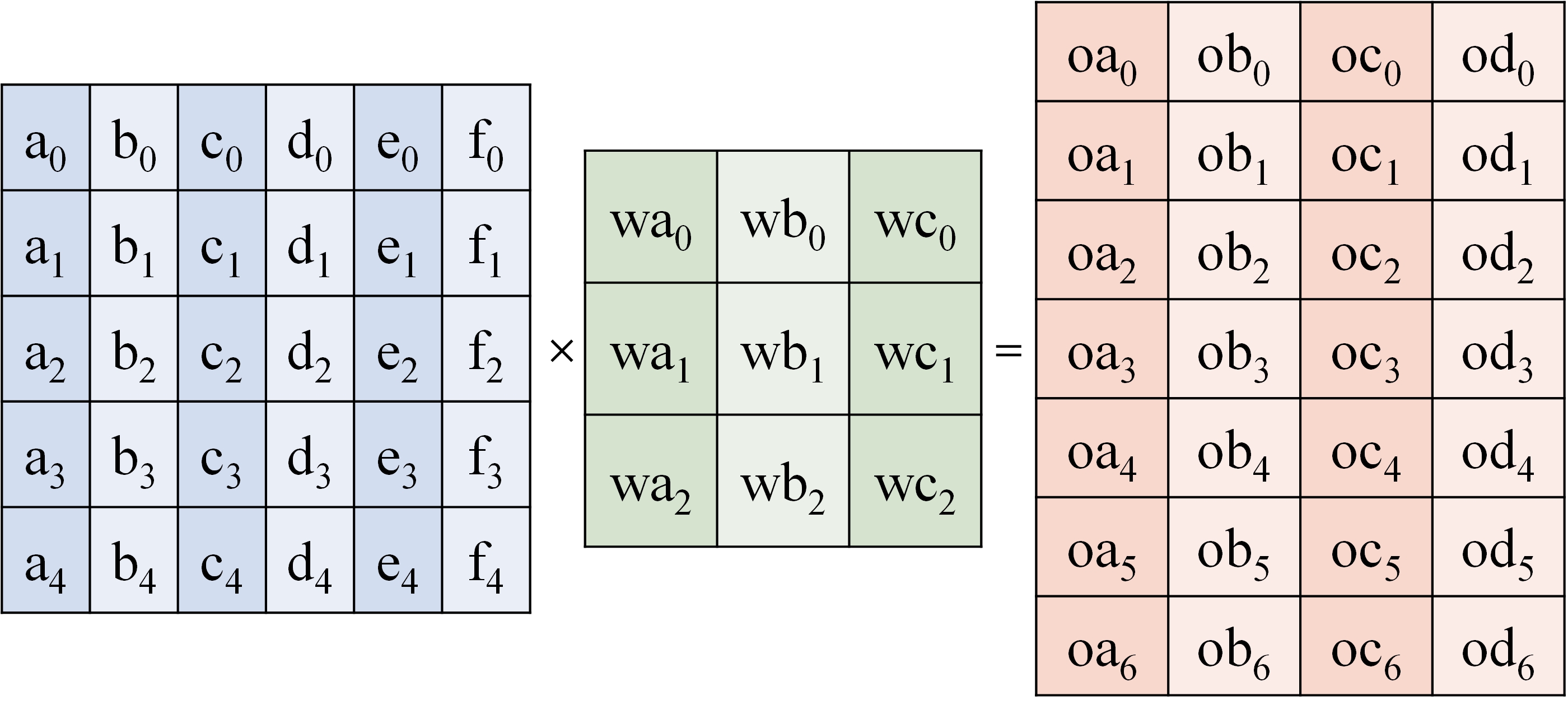}}
	\caption{\textcolor{black}{A convolution example with 5$\times$6 input, 3$\times$3 weight, and 7$\times$4 output.}}
	\label{fig.example}
\end{figure}
\begin{figure*}[t]
	\centering{\includegraphics[height=70mm]{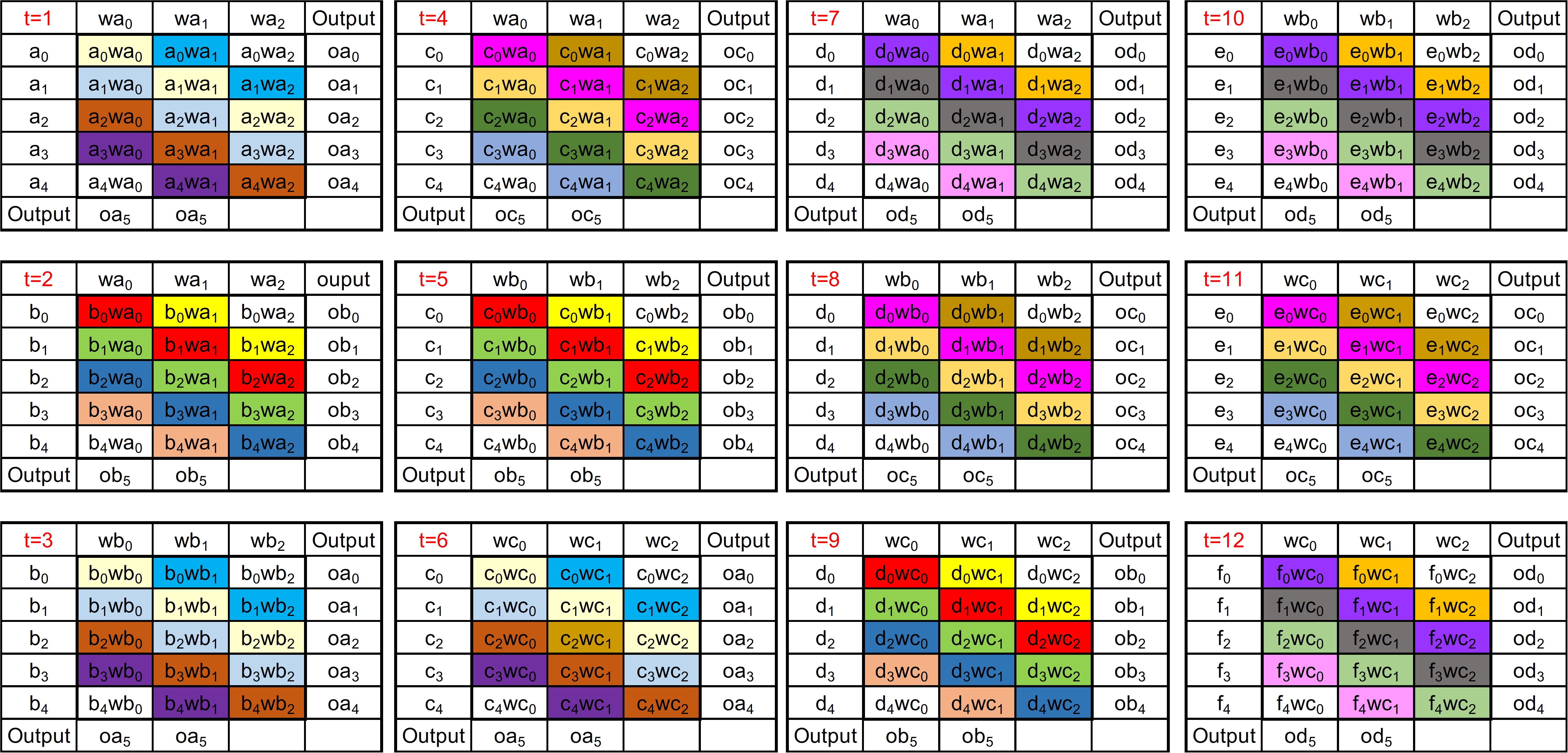}}
	\caption{\textcolor{black}{Dataflow chart for the example in Fig.~\ref{fig.example} with 5x3 MACs. In which, the same color elements belong to the same filter output, which will be summed together in a PE at the same cycle and then accumulated later in the accumulator at different cycles, e.g. black elements at t = 2, 5, and 9 accumulated for $ob_2$.}}
	\label{fig.dataflow}
\end{figure*}

\begin{figure}[t]
	\centering{\includegraphics[height=40mm]{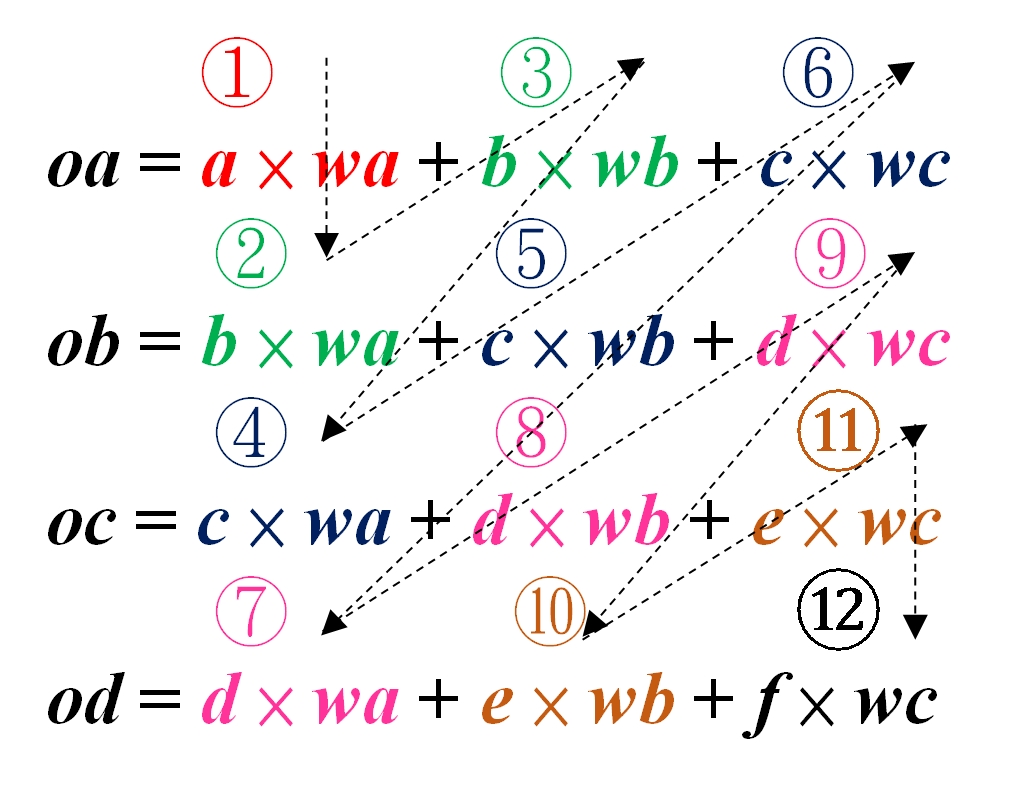}}
	\caption{The computational order for 3$\times$3 convolutions with unit stride, \textcolor{black}{where $\boldsymbol{oa}$ to $\boldsymbol{od}$ are the column vector of the output. $\boldsymbol{a}$ to $\boldsymbol{f}$ are the column vector of the input. $\boldsymbol{wa}$ to $\boldsymbol{wc}$ are the column vector of the weight.} Numbers 1-12 represent cycle numbers in Fig.~\ref{fig.dataflow}. }
	\label{fig.3x3}
\end{figure}
\begin{figure}[t]
	\centering{\includegraphics[height=25mm]{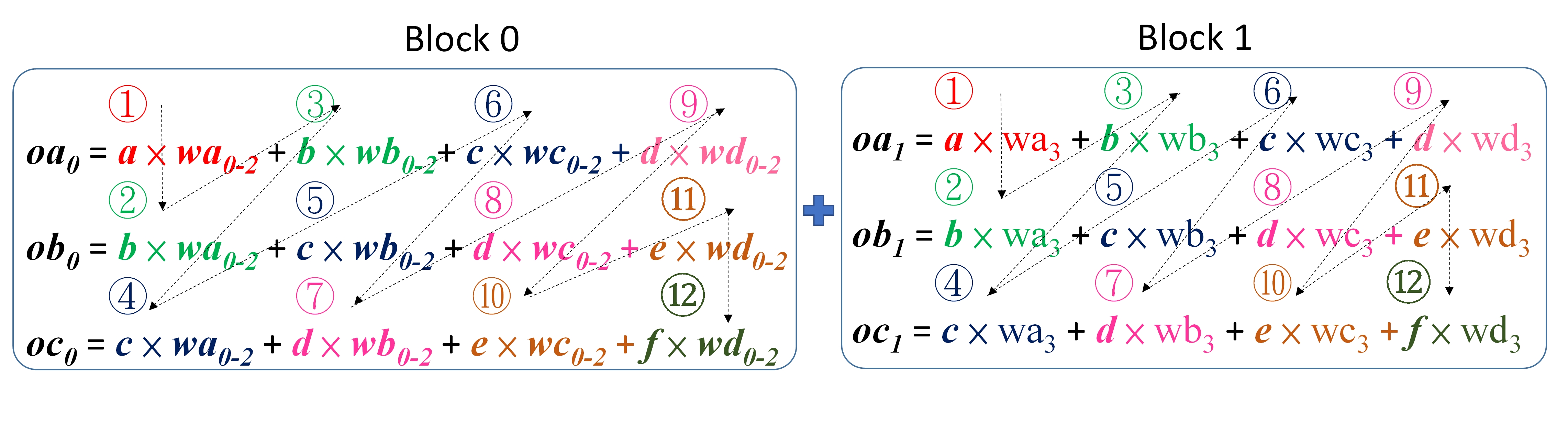}}
	\caption{The computational order for 4$\times$4 convolutions with unit stride, \textcolor{black}{where $\boldsymbol{wa_0-_2}$ is a vector with $wa_0$, $wa_1$, $wa_2$,  $\boldsymbol{wb_0-_2}$ is a vector with $wb_0$, $wb_1$, $wb_2$, and so on. $\boldsymbol{oa} = \boldsymbol{oa_0} + \boldsymbol{oa_1}$, $\boldsymbol{ob} = \boldsymbol{ob_0} + \boldsymbol{ob_1}$, $\boldsymbol{oc} = \boldsymbol{oc_0} + \boldsymbol{oc_1}$.}}
	\label{fig.4x4}
\end{figure}
\begin{figure}[t]
	\centering{\includegraphics[height=45mm]{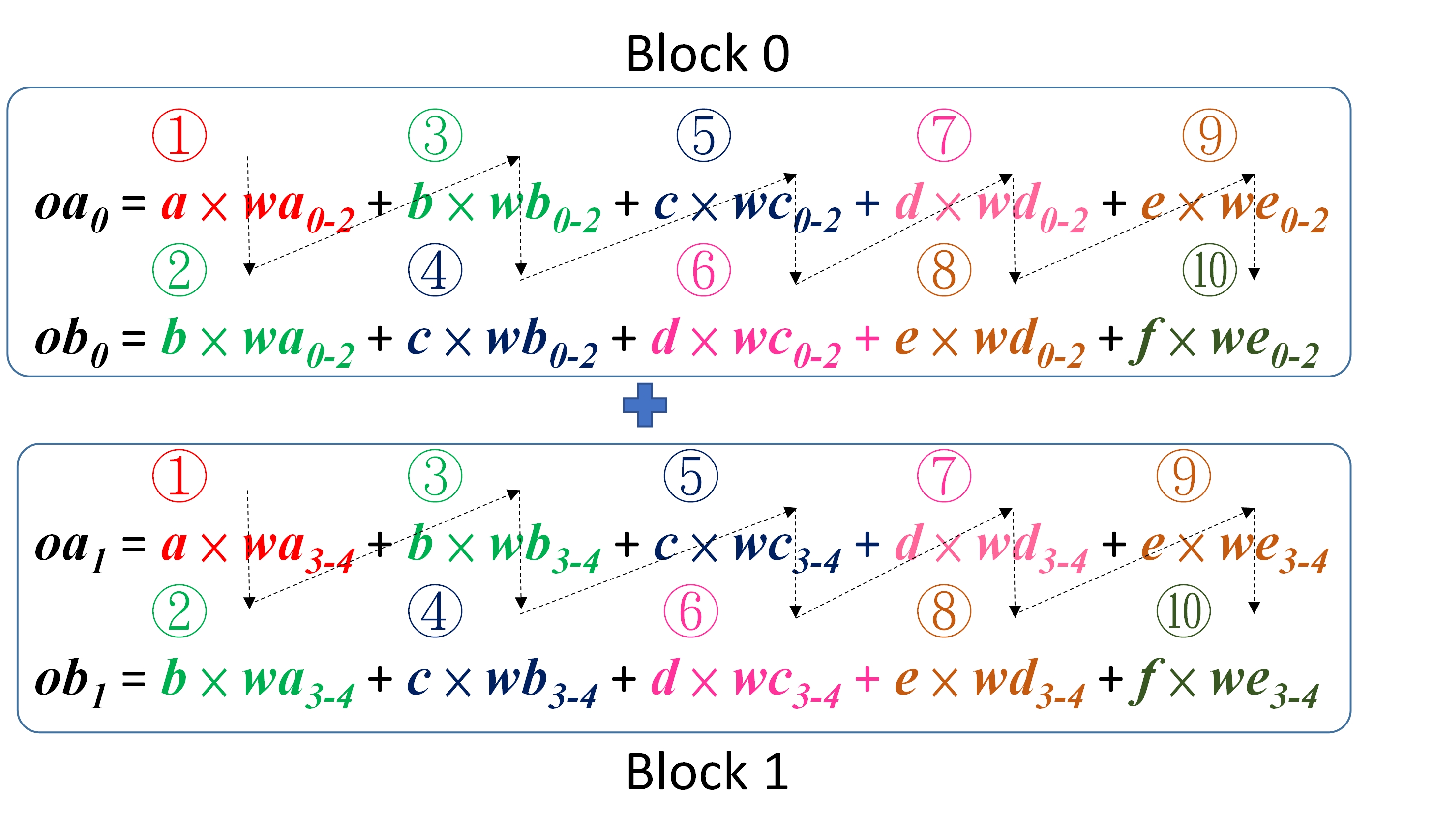}}
	\caption{The computational order for 5$\times$5 convolutions with unit stride, \textcolor{black}{where $\boldsymbol{wa_3-_4}$ is a vector with $wa_3$, $wa_4$, $\boldsymbol{wb_3-_4}$ is a vector with $wb_3$, $wb_4$, and so on. $\boldsymbol{oa} = \boldsymbol{oa_0} + \boldsymbol{oa_1}$, $\boldsymbol{ob} = \boldsymbol{ob_0} + \boldsymbol{ob_1}$.}}
	\label{fig.5x5}
\end{figure}
\begin{figure}[t]
	\centering{\includegraphics[height=35mm]{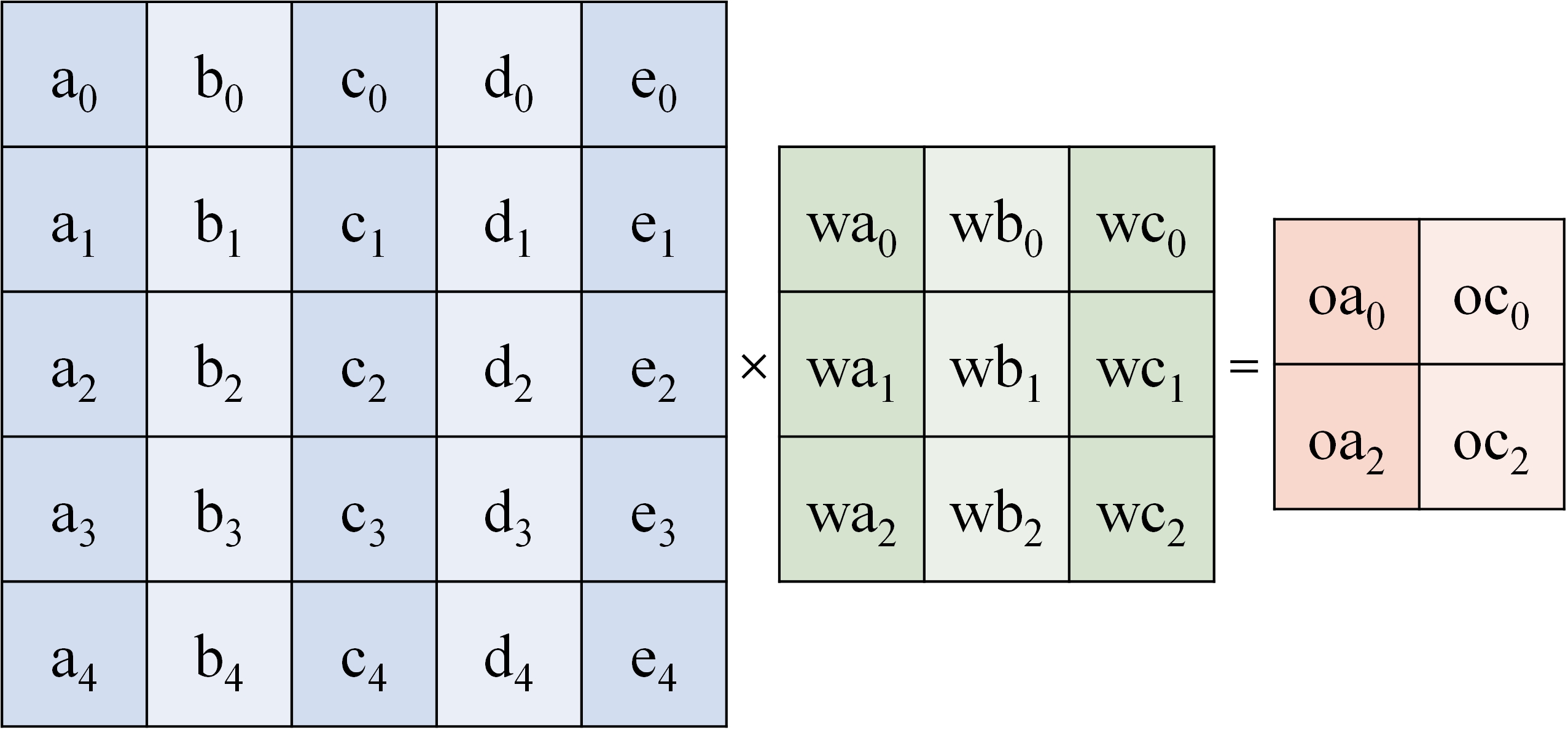}}
	\caption{\textcolor{black}{An example of 3x3 convolutions/stride 2 with 5×5 input to generate 2×2 output.}}
	\label{fig.examples2}
\end{figure}
\begin{figure*}[t]
	\centering
	\subfigure{\includegraphics[height=38mm]{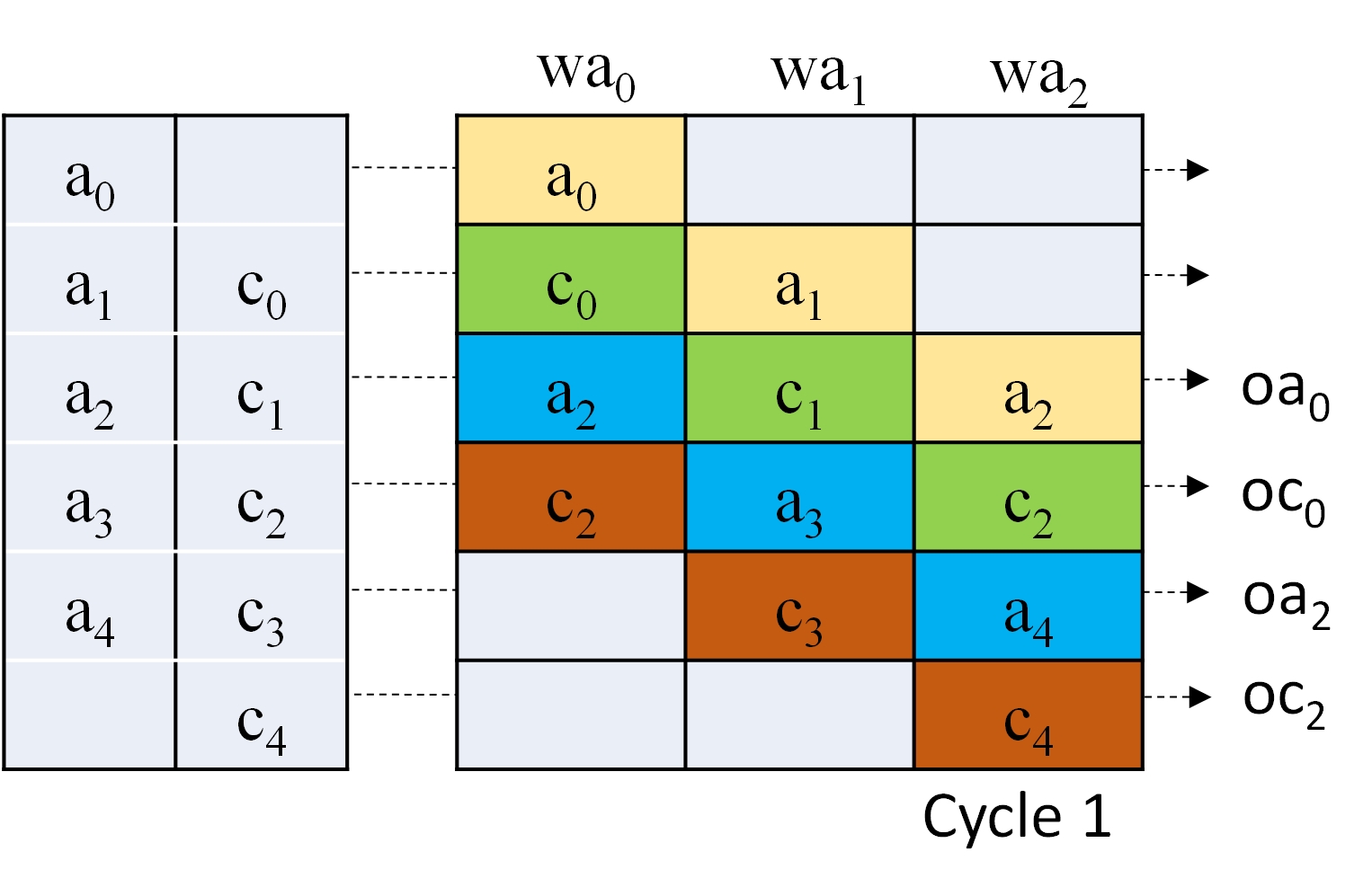}}
	\subfigure{\includegraphics[height=38mm]{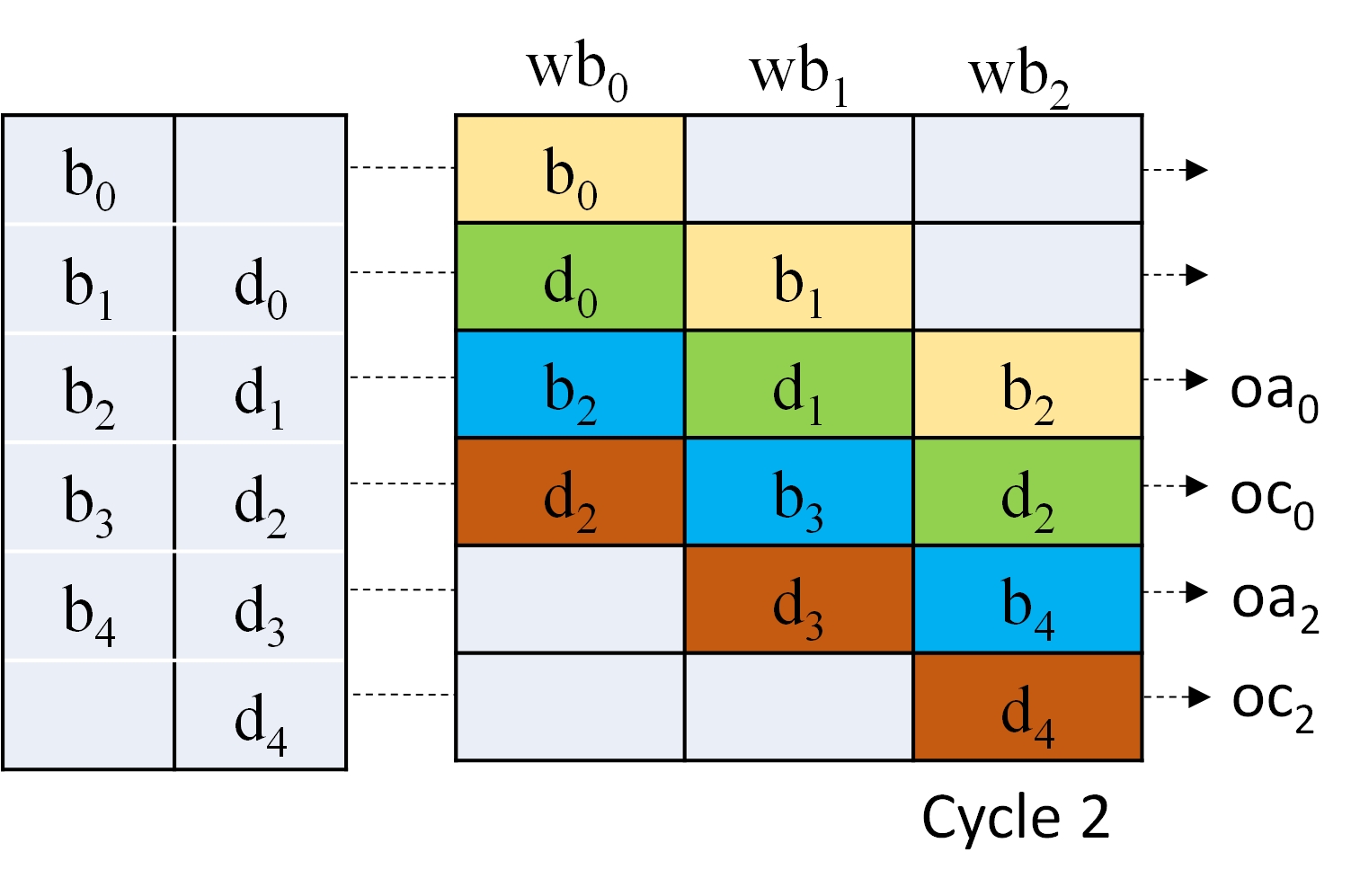}}
	\subfigure{\includegraphics[height=38mm]{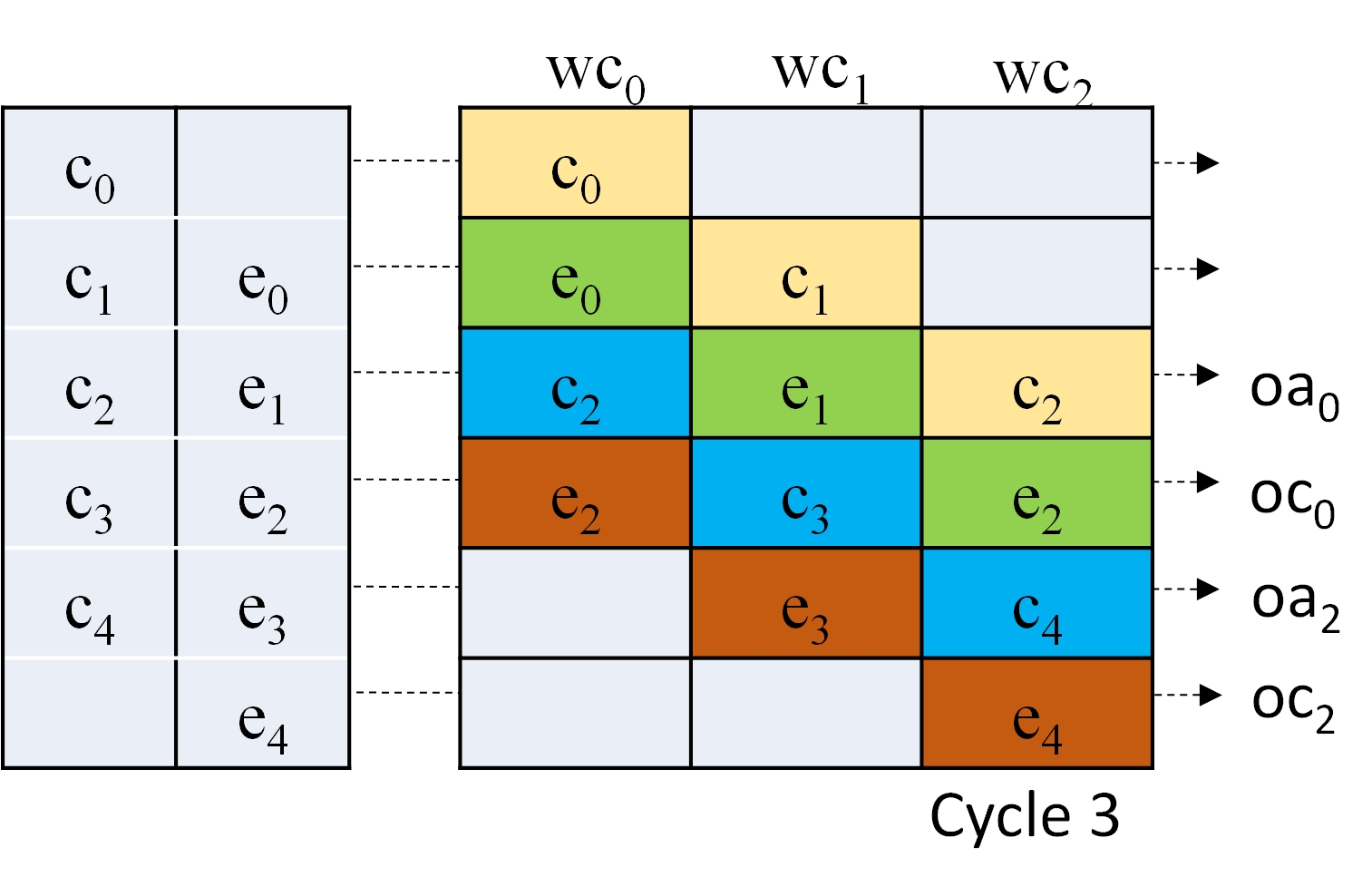}}
	\caption{Flow chart for 3$\times$3 convolutions with stride 2 \textcolor{black}{example in Fig.~\ref{fig.examples2}.}}
	\label{fig7:3x3s2}
\end{figure*}

\begin{figure}[t]
	\centering
	\subfigure{\includegraphics[height=35mm]{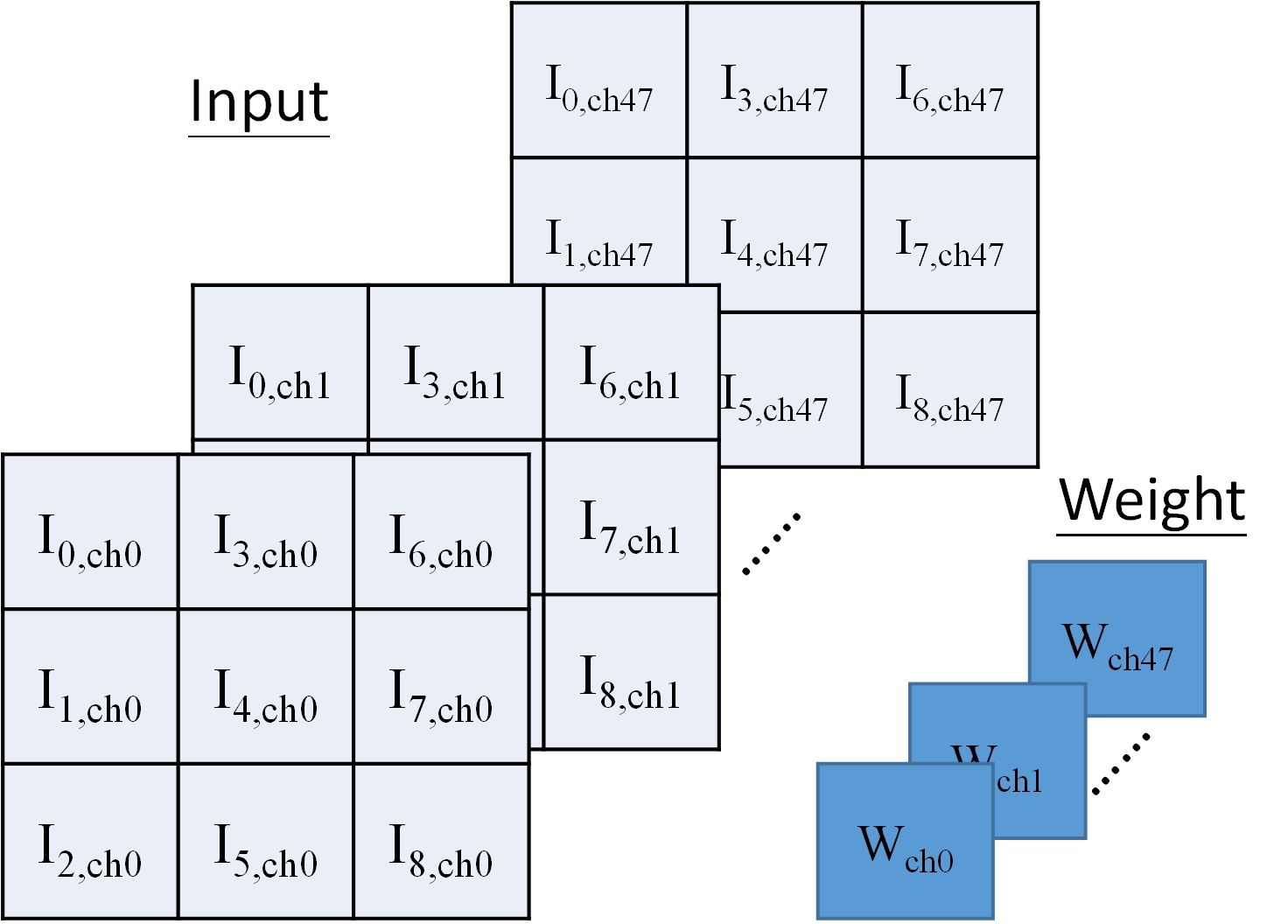}}
	\subfigure{\includegraphics[height=80mm]{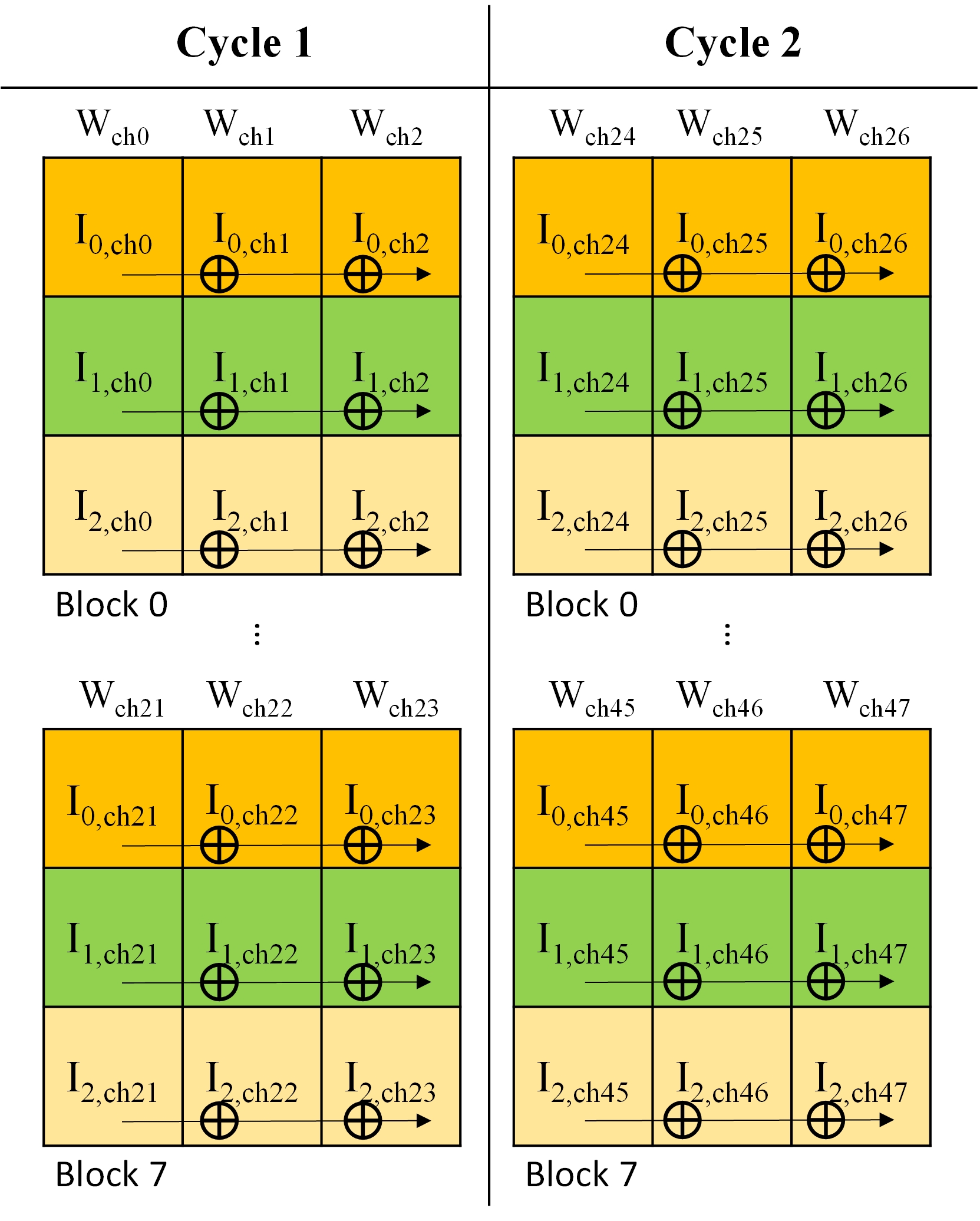}}
	\caption{Flow chart for 1$\times$1 convolutions. \textcolor{black}{The same color elements will be summed together.}}
	\label{fig:1x1}
\end{figure}
\begin{figure}[t]
	\centering
	{\includegraphics[height=80mm]{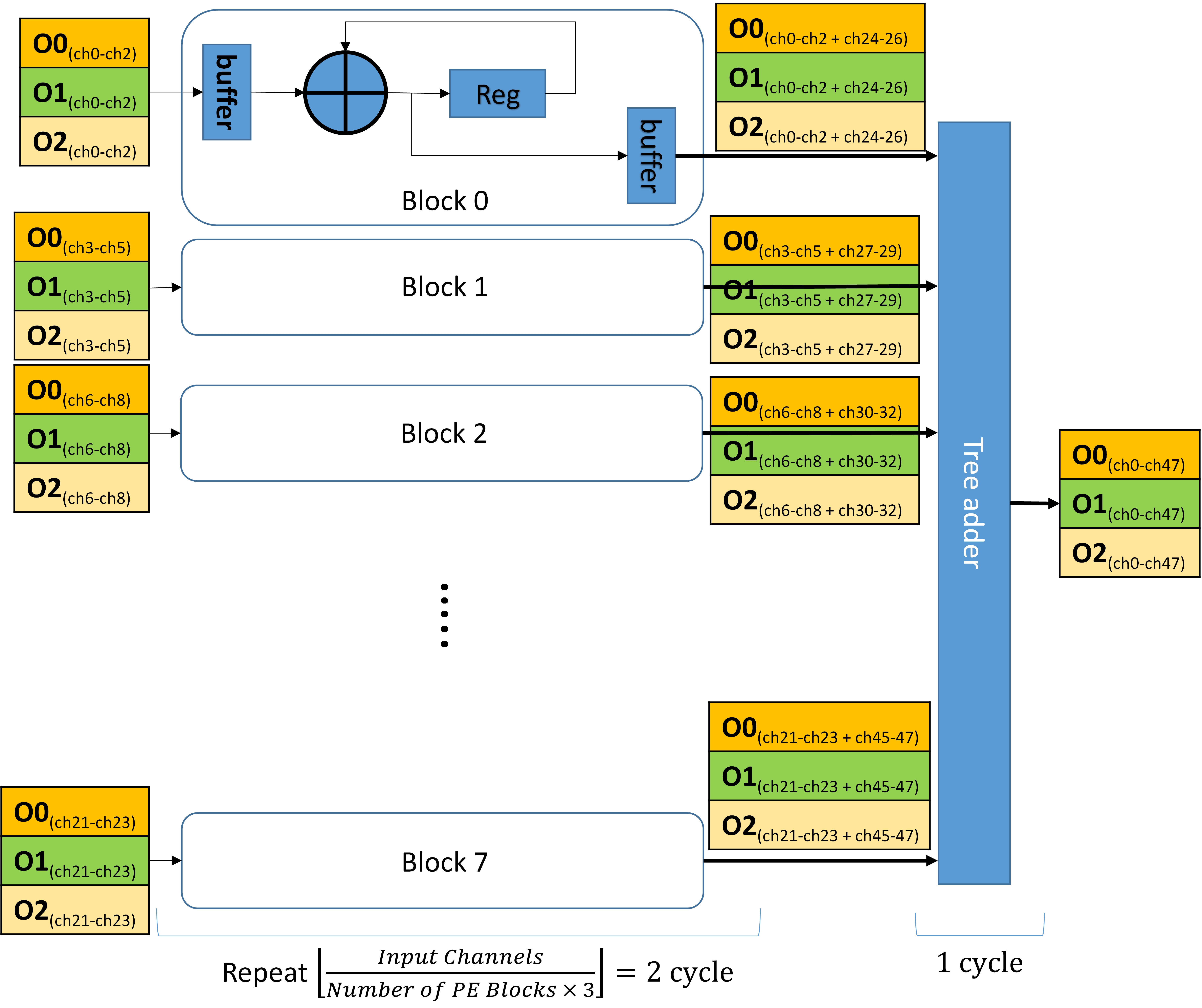}}
	\caption{\textcolor{black}{Accumulation diagram for 1$\times$1 convolutions. $\mathrm{O0_(ch0-ch2)}$ represents that one output pixel have been accumulated with input channels from 0 to 2 in the stage.}}
	\label{fig:1x1_acc}
\end{figure} 
\section{Data Flow of Different Kernels}
The proposed accelerator is designed to optimize computations of 3$\times$3 filters stride one  with a vectorwise scheduling due to its wide usage. For larger filter size, we adopt kernel decomposition method \cite{yj} so that this design only needs extra support of 4$\times$4, and 5$\times$5 filters. In addition, with the popularity of low complexity models, this design also supports 3$\times$3 with stride two by interleaved input, depthwise convolution by reconfigurable accumulators and 1$\times$1 filter by elementwise input dataflow for higher hardware utilization.

\subsection{\textcolor{black}{Basic 3$\times$3 Convolution with Unit Stride}}
Fig.~\ref{fig.dataflow} shows the data flow of a PE block to compute \textcolor{black}{an example with} 5$\times$6 input tile and a 3$\times$3 filer kernel as in Fig.~\ref{fig.example}. This example case assumes a 5$\times$3 MAC array for clarity. In which, the blocks with the same color are the partial results for the same convolution output. Above computation also includes tile boundary cases on top \textcolor{black}{(e.g. $oa_0$ - $oa_1$, $ob_0$ - $ob_1$, $oc_0$ - $oc_1$, $od_0$ - $od_1$)} and bottom (e.g. $oa_5$ - $oa_6$, $ob_5$ - $ob_6$, $oc_5$ - $oc_6$, $od_5$ - $od_6$), which will be stoblack in the global boundary buffer. For a 3$\times$3 convolution, one column vector of input is broadcasted horizontally along the MAC row (e.g. $a_0$ - $a_4$ when t = 1) and one column vector of filter weight is broadcasted vertically (e.g. $wa_0$ - $wa_2$ when t = 1). With this broadcasted data flow, the multiplication result along the diagonal direction will belong to the same output pixels (e.g. $oa_0$ - $oa_6$ when t=1), which will be summed together in the same cycle as the multiplications occur, and then accumulated sequentially in the following accumulator. This vectorwise scheduling enables simple dataflow for high hardware utilization and low area cost. For 3$\times$3 depthwise convolution, the output of the first stage of the accumulators will be the desiblack result. 

The overall vectorwise scheduling is shown in Fig.~\ref{fig.dataflow}, which takes 12 cycles to complete the example case. The concept of this scheduling is to decompose a convolution into sums of \textcolor{black}{MAC results by} an input column and a weight column as shown in Fig.~\ref{fig.3x3}. Each MAC result of an input column and a weight column will take one cycle to complete. Thus, a 3$\times$3 filter computation will need three cycles. However, \textcolor{black}{these three cycles are not continuous}. We rearrange the computation order as shown in Fig.~\ref{fig.3x3} so that the same input can be reused successively for the computation to save the power of input buffer SRAM. Besides, this reorder enables a simple continuous addressing of the input SRAM buffer to blackuce control overhead. 

With the above computation, the speedup \textcolor{black}{for the above example} is $(number~of~MACs)~/~cycles = 180/12 = 15$, which is equal to the number of PEs $(3\times5=15)$. Thus, the hardware is fully utilized.


\subsection{\textcolor{black}{4$\times$4 and 5$\times$5 Convolution with Unit Stride}}
Beyond 3$\times$3 convolution with unit stride, our proposed architecture also supports 4$\times$4 and 5$\times$5 convolutions to implement different filter sizes in current popular models. The 4$\times$4 and 5$\times$5 convolutions will take two PE blocks.  As shown in Fig.~\ref{fig.4x4} and Fig.~\ref{fig.5x5}, the first three elements \textcolor{black}{(e.g. $wa_0$ - $wa_2$, $wd_0$ - $wd_2$ in 4$\times$4, or $wa_0$ - $wa_2$, $we_0$ - $we_2$ in 5$\times$5)} of the filter columns will be at the first block and the remaining elements \textcolor{black}{(e.g. $wa_3$, $wd_3$ in 4$\times$4, or $wa_3$ - $wa_4$, $we_3$ - $we_3$ in 5$\times$5)} will be at the second block with two or one column of unused MACs. The overall data flow is the same as the one in the 3$\times$3 convolution but with properly weight assignments for 4$\times$4 and 5$\times$5. With this approach, the hardware utilization will be lower since part of the PE block will not be used. This design tradeoff can make hardware regular since these kernel sizes are only used in the first input layer in most of the current network models.

\subsection{\textcolor{black}{3$\times$3 Convolution with Stride 2}}
For a 3$\times$3 convolution with stride 2, a naive dataflow implies a 25\% hardware utilization since only 25\% of PE result is useful for the final result. To increase hardware utilization, we propose interleaved input as shown in Fig.~\ref{fig7:3x3s2}. Thus, we add a multiplexer at the input data of multiplier in each PE as shown in Fig.~\ref{fig:PE} to support this. Fig.~\ref{fig7:3x3s2} shows the data flow of a PE block for \textcolor{black}{an example with} a 5$\times$5 input, and 3$\times$3 filer kernel with stride 2. Input data from two different input columns (e.g. $a_0$ - $a_4$, $c_0$ - $c_4$) will be interleaved as the input of successive PE rows to increase hardware utilization. An example as shown in Fig.~\ref{fig.examples2} with stride 2 only need 3 cycles to generate $oa_0$, $oa_2$, $oc_0$, and $oc_2$. The result will be accumulated at the accumulator as 3$\times$3 convolution with unit stride but with interleaved results. Thus, the speedup is $(number~of~ MACs)/cycles = 36/3 = 12$. However, the PE block \textcolor{black}{will have some MACs unused due to} interleaved input. So the utilization is $speedup / number~of~PEs = 12/18 = 67\%$ in this example, which is lower than \textcolor{black}{that in} 3$\times$3 convolutions with unit stride but higher than \textcolor{black}{that with a} naive mapping.
\begin{figure}[t]
	\centering
	{\includegraphics[height=50mm]{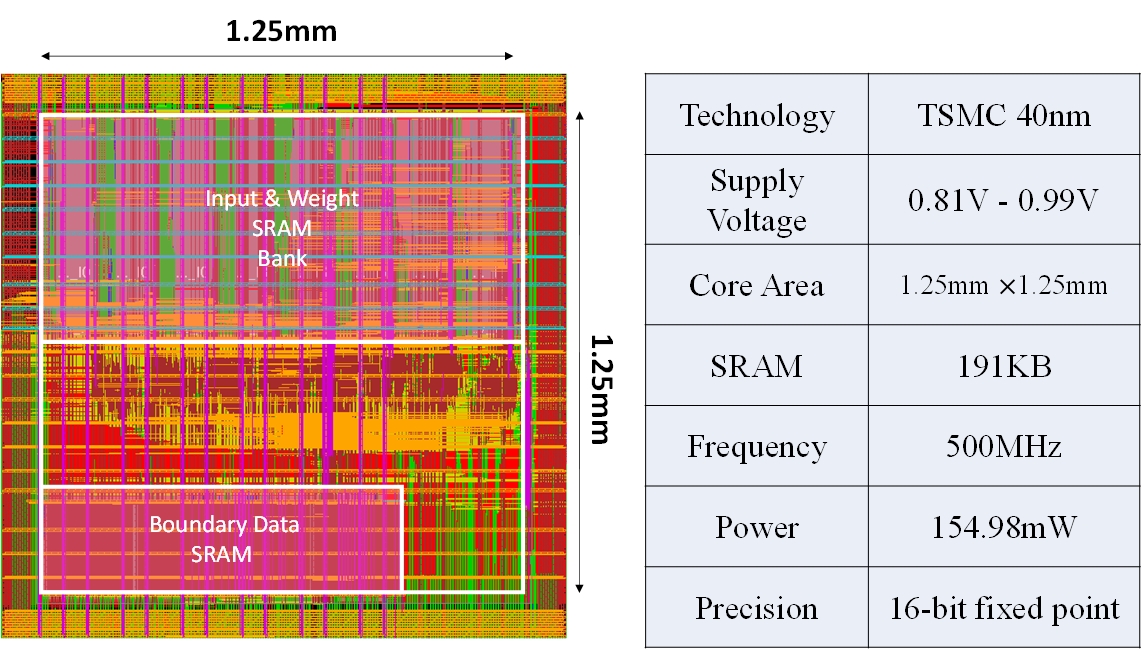}}
	\caption{Implementation result.}
	\label{fig.layout}
\end{figure} 
\begin{table}[t]
	\centering
	\caption{\textcolor{black}{Per layer DRAM access amount for VGG-16.}}
	\label{table:energy}
	\begin{tabular}{|l|c|c|c|c|c||c|}
		\hline
		
		\multirow{3}{0.5cm}{Layers}& Input  & Weight & Output  & Total  & Total&Tile  \\
		&Access&Access&Access&Access&Energy& Size \\
		&(MB)&(MB)&(MB)&(MB)& (mJ)&(7$\times$n)\\\hline
		1 &0.287 & 0.002& 6.125& 6.432& 0.45&224\\\hline
		2 &6.125 &0.035&6.125 & 12.285& 0.86 &112\\\hline
		3 & 1.531& 2.25& 3.063& 6.844&  0.479&112\\\hline
		4 & 3.063& 9& 3.063&15.125&1.059 &56\\\hline
		5 & 0.766& 4.5& 1.531& 6.797& 0.476&56\\\hline
		6 & 1.531& 18& 1.531& 21.063& 1.474&56\\\hline
		7 & 1.531& 18& 1.531& 21.063& 1.474&56\\\hline
		8 & 0.383&9& 0.766& 10.148& 3.02&28\\\hline
		9 & 0.766& 36& 0.766& 37.531& 2.627&14\\\hline
		10 & 0.766& 36& 0.766& 37.531& 2.627&14\\\hline
		11 & 0.191& 9& 0.191& 9.383& 0.657&14 \\\hline
	    12 & 0.191& 9& 0.191& 9.383& 0.657&14 \\\hline
	    13 & 0.191& 9& 0.191& 9.383& 0.657&14 \\\hline
		Total & 17.322 &159.805 &25.84 &202.967 &14.208&\\\hline
	\end{tabular}
\end{table}

\subsection{\textcolor{black}{1$\times$1 Convolution}}
 1$\times$1 convolutions are getting popular in modern deep learning models. However, unlike other convolutions, the 1$\times$1 convolution only has size one in the spatial dimension, which will result in much more output than 3$\times$3 convolutions if we implement it naively on the current 3$\times$3 convolution based PE blocks. This implies more parallel hardware needed for the following accumulators to keep PE array full utilized or the PE array \textcolor{black}{stalled} during the accumulation. 

To solve above problems, we propose the elementwise input scheduling for 1$\times$1 convolutions as shown in Fig.~\ref{fig:1x1} that will have distinctive input elements in each multiplier. In this scheduling, each PE block will select their specific input feature map via a \textcolor{black}{3-to-1} multiplexer. With such input, the summation is now along the horizontal direction. Thus, we will have output number equal to $ block~number = 8 $ for eight PE blocks. Each output from a PE block will have sum of three channel multiplications. In this flow, all PE outputs are just different parts of a channel output. Thus, we will accumulate them with the first two stages of the accumulator as shown in Fig.~\ref{fig:1x1_acc}. The stage three is skipped in this case. With above flow, we can achieve high hardware utilization.
\begin{figure}[t]
	\centering
	{\includegraphics[height=50mm]{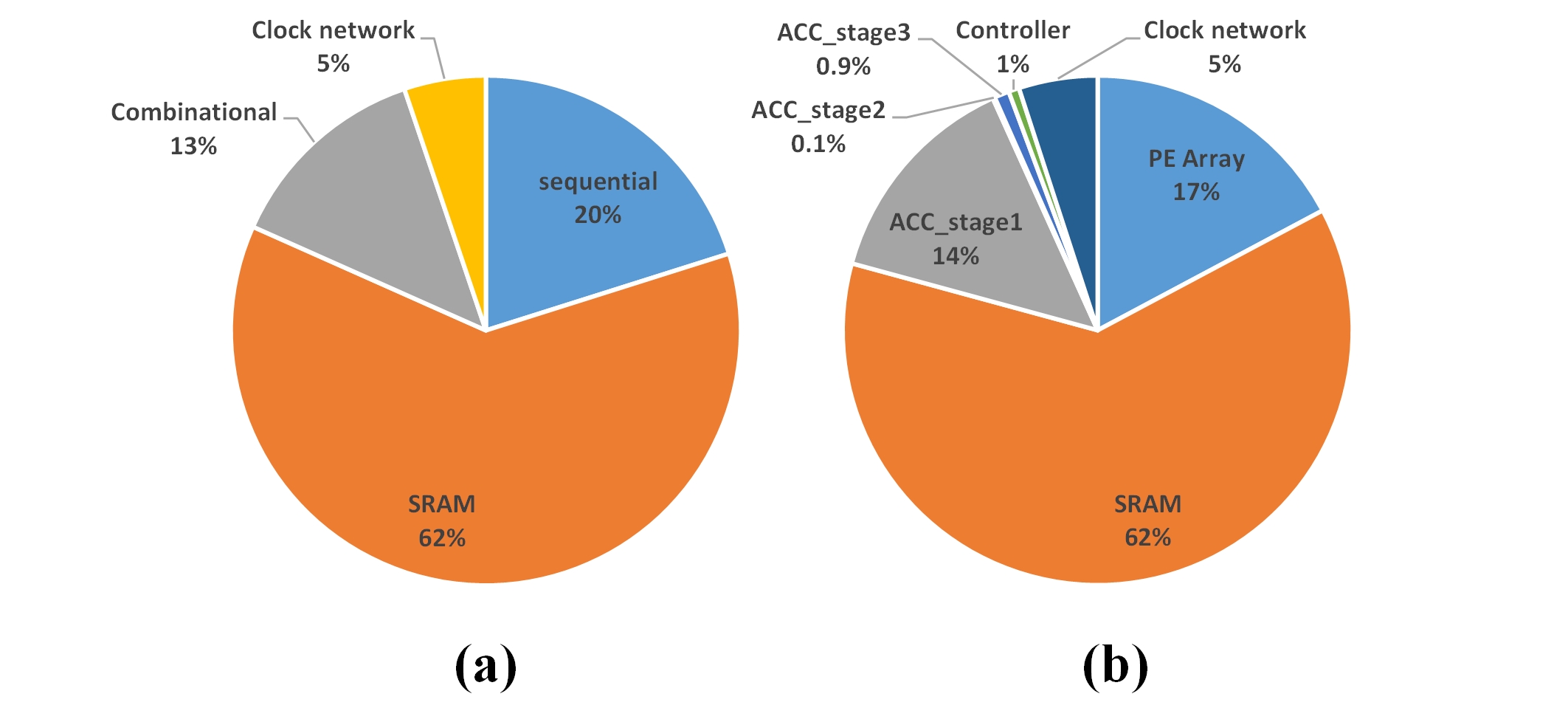}}
	\caption{\textcolor{black}{Power breakdown of (a) logic circuits, (b) architecture elements in the core.}}
	\label{fig.power}
\end{figure} 
\begin{figure}[t]
	\centering
	{\includegraphics[height=50mm]{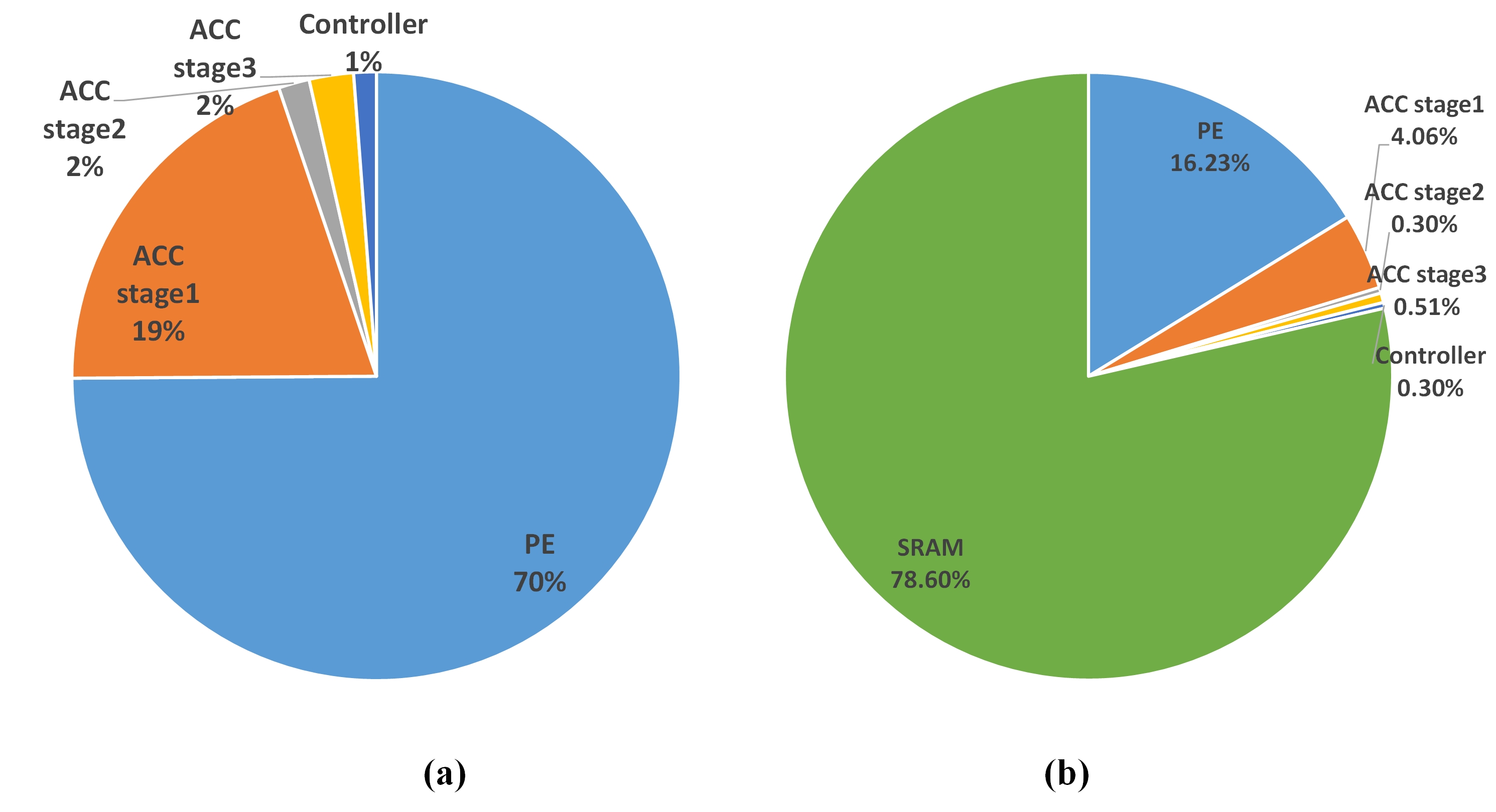}}
	\caption{(a) Area breakdown of the core without SRAM, (b) with SRAM. }
	\label{fig.area}
\end{figure}

\begin{figure}[t]
	\centering
	{\includegraphics[height=50mm]{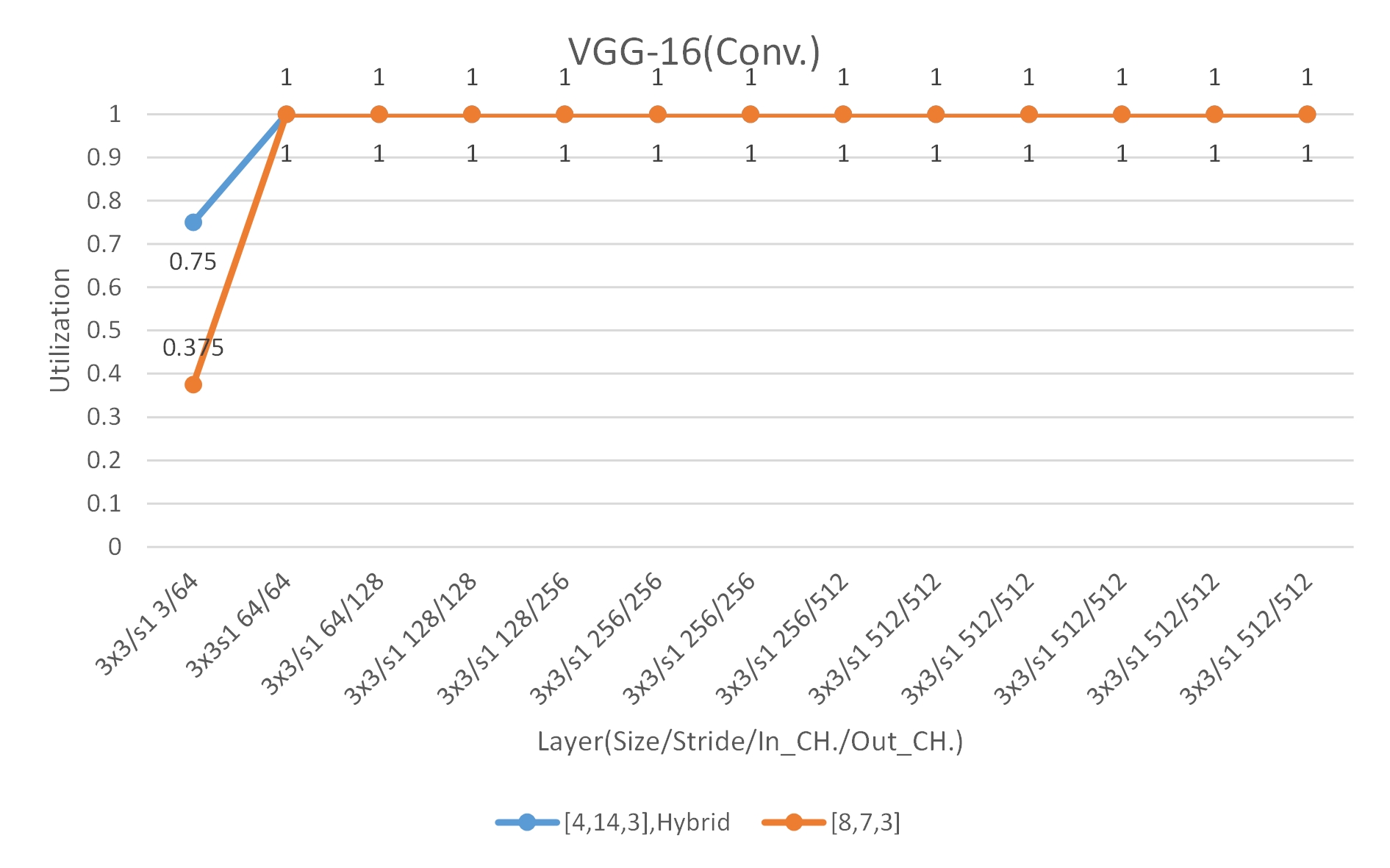}}
	\caption{Per layer hardware utilization of the VGG-16.}
	\label{fig:vgg_u}
\end{figure}

\begin{figure}[t]
	\centering
	{\includegraphics[height=50mm]{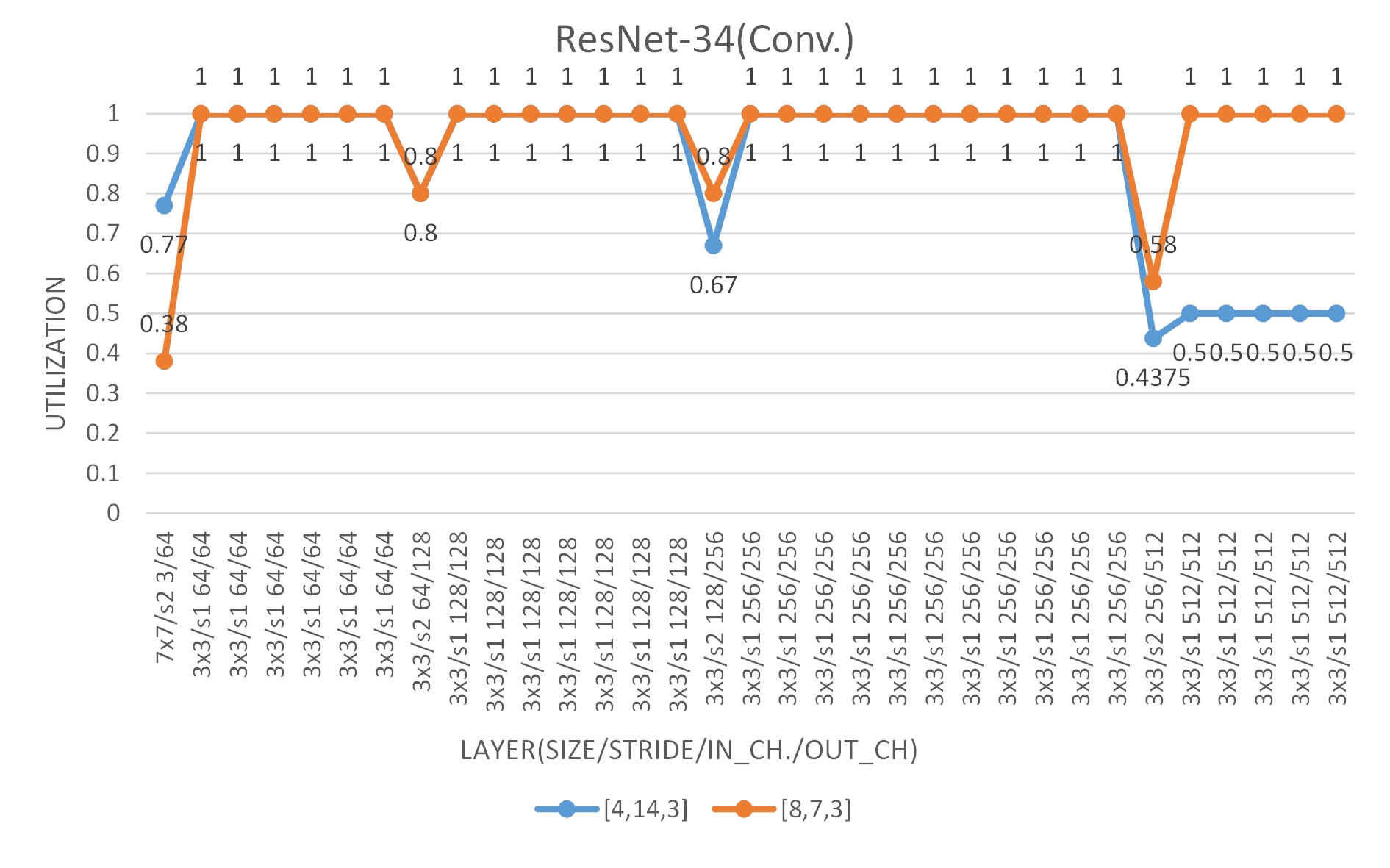}}
	\caption{Per layer hardware utilization of the ResNet-34.}
	\label{fig:res_u}
\end{figure}

\begin{figure}[t]
	\centering
	{\includegraphics[height=50mm]{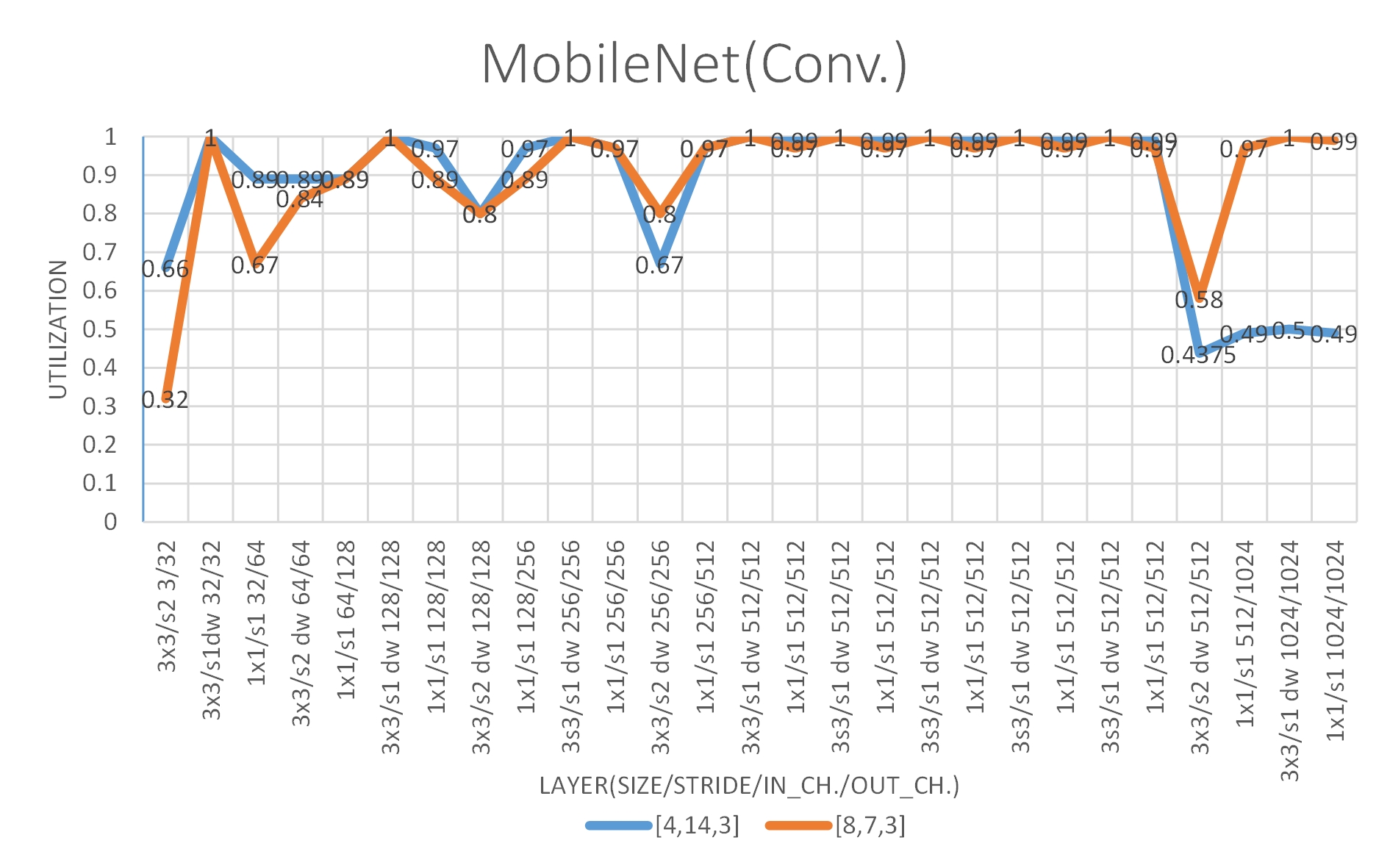}}
	\caption{Per layer hardware utilization of the MobileNet.}
	\label{fig:mob_u}
\end{figure}

\begin{figure}[t]
	\centering
	{\includegraphics[height=50mm]{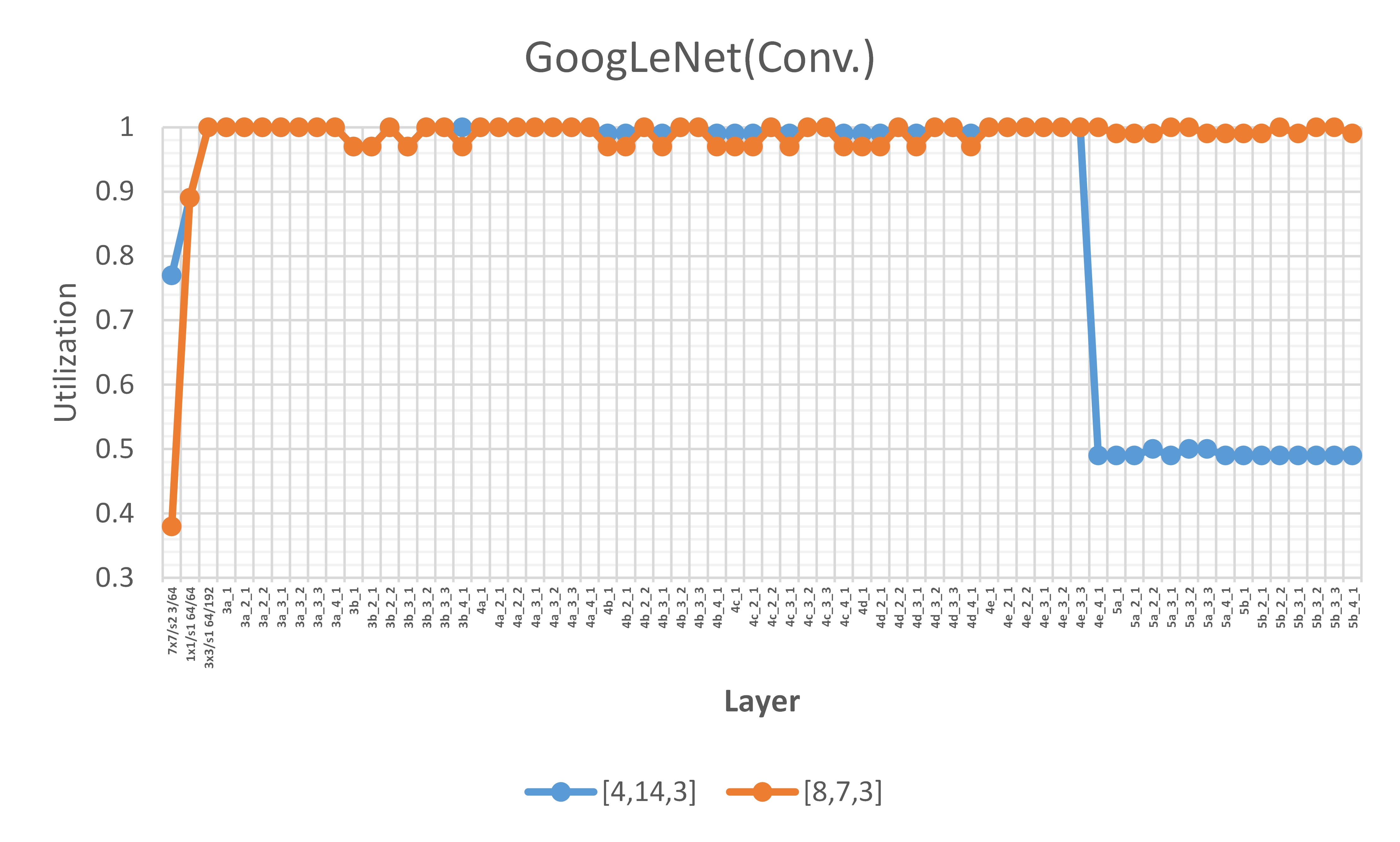}}
	\caption{\textcolor{black}{Per layer hardware utilization of the GoogLeNet. In which, 3a\_2\_2 represents Incepetion 3a, second branch, and second layer.}}
	\label{fig:google}
\end{figure}
\section{Experimental Result}
\subsection{Experiment Setup}
This design has been implemented with the TSMC 40nm CMOS technology. The hardware power consumption is estimated by Synopsys PrimeTime PX. We also build a cycle based tool by Pytorch to properly evaluate hardware utilization, speedup and DRAM access under different models and PE array configurations. In addition, this tool also helps generate configuration information to configure the hardware controller. For the following analysis, we will use four famous models, VGG-16\cite{VGG}, MobileNet\cite{mobilenet}, ResNet-34\cite{resnet}, and GoogLeNet\cite{googlenet}, trained on the ImageNet dataset as our test cases.

\subsection{\textcolor{black}{Implementation Result}}
Fig.~\ref{fig.layout} shows the chip layout and its implementation result. The core area is 1.25mm$\times$1.25mm with 191KB SRAM including 99KB input SRAM bank, 36KB weight SRAM bank, and 56KB global boundary data SRAM. The peak performance is 168 GOPS at 500MHz based on \textcolor{black}{the technology library from the worst PVT corner}, when the PE utilization is 100\%. The power consumption is 154.98mW, measublack by running VGG-16 convolutions \textcolor{black}{based on the technology library from the best PVT corner at 0.99 V power supply.}

\subsection{Memory Access and Energy Consumption Analysis}
The external memory access of this design is blackuced by data reuse. The input is divided into configurable tile size with 7$\times$n size \textcolor{black}{for the ImageNet dataset, where} n is denoted as in Table~\ref{table:energy}. The input SRAM buffer has size enough to store 7$\times$6 tiles with 1024 input channels for commonly used models. The proposed design will first compute all channels of an input tile with all weight data to get the final output tile. During this computation, all input data, partial sum, and output will access local buffer instead of external DRAM. Only the weight data will be accessed from DRAM repeatedly when the input tiles are changed. So, external DRAM access can be formulated as follows:
\begin{equation}
Input~Access = input~channel \times image~size
\end{equation}
\begin{equation}
\begin{split}
Weight~Access=number~of~input~tiles\times \\output~channel\times input~channel \times weight~size
\end{split}
\end{equation}
\begin{equation}
Output~Access = output~channel \times output~size
\end{equation}

The total energy is $70pJ/bit \times total~access$ for DDR3 DRAM\cite{power}. As shown in Table.~\ref{table:energy}, there are no blackundant accesses in input and output. The weight accesses are correlated with the number of tiles. With the configurable tile size, we can tradeoff between tile size and input channel to decrease the number of tiles and weight accesses.

Fig.~\ref{fig.power}(a) shows power breakdown of the logic circuit in the core. SRAM accounts for 62\% \textcolor{black}{of the total power}. The combinational logic consumes 13\% for MAC and accumulator computations. The sequential logic consumes only 20\%, because of the lower register numbers in this data broadcasted design. This also results in lower power consumption in the clock network. \textcolor{black}{Fig.~\ref{fig.power}(b) shows power breakdown of architecture elements in the core. The PE array accounts for 17\% due to 168 MACs. The stage one of the accumulator occupies 14\% because it keeps accumulation in most of the cycles and consists of eight parallel accumulators. The other stages of the accumulator occupies only 1\% because they only operate after a valid stage one output.}

\begin{figure*}[t]
	\centering
	\subfigure[Utilization in different models and PE configurations.]{\includegraphics[height=55mm]{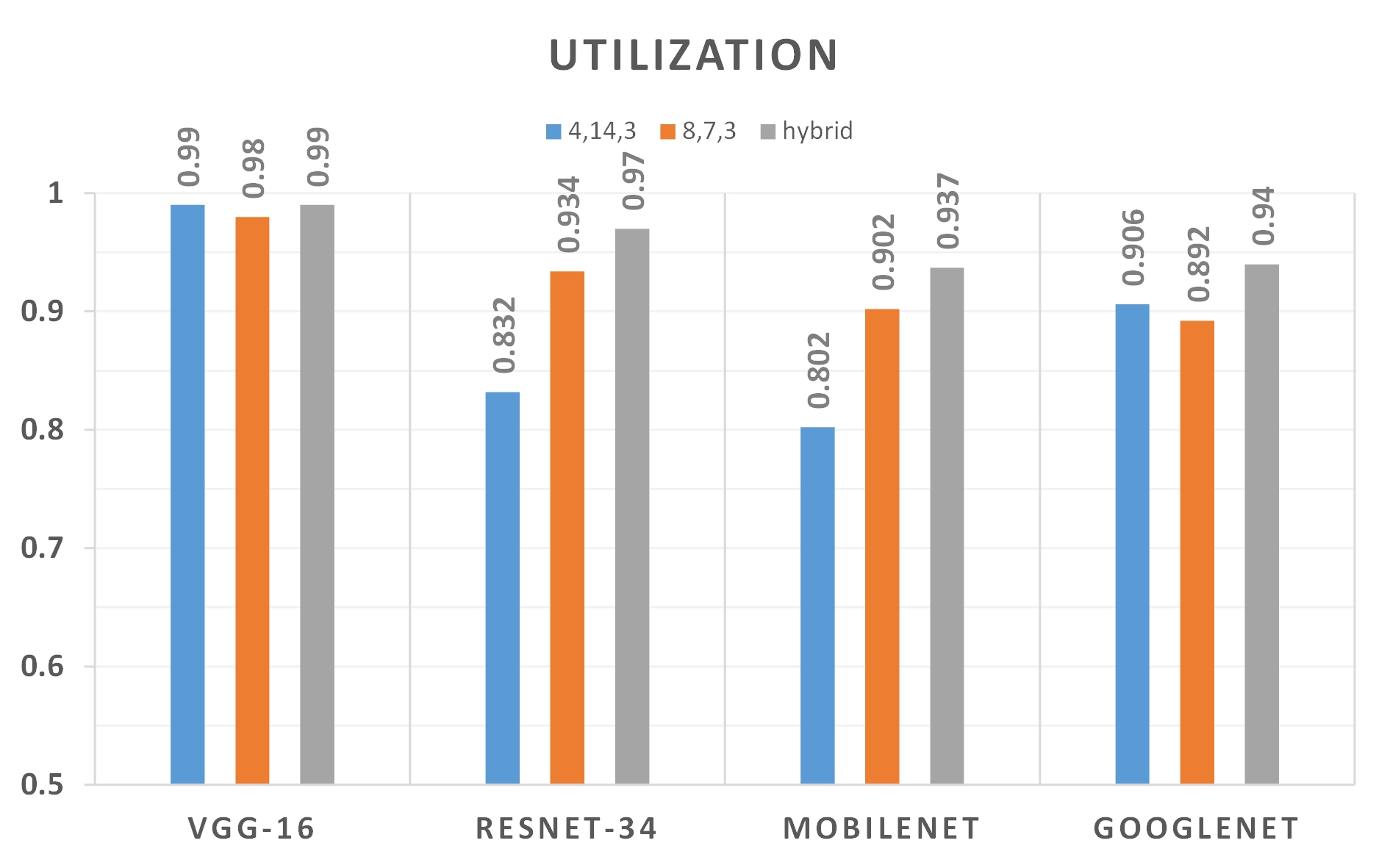}\label{fig:utilization}}
	\subfigure[Throughput in different models and PE configuration.]{\includegraphics[height=55mm]{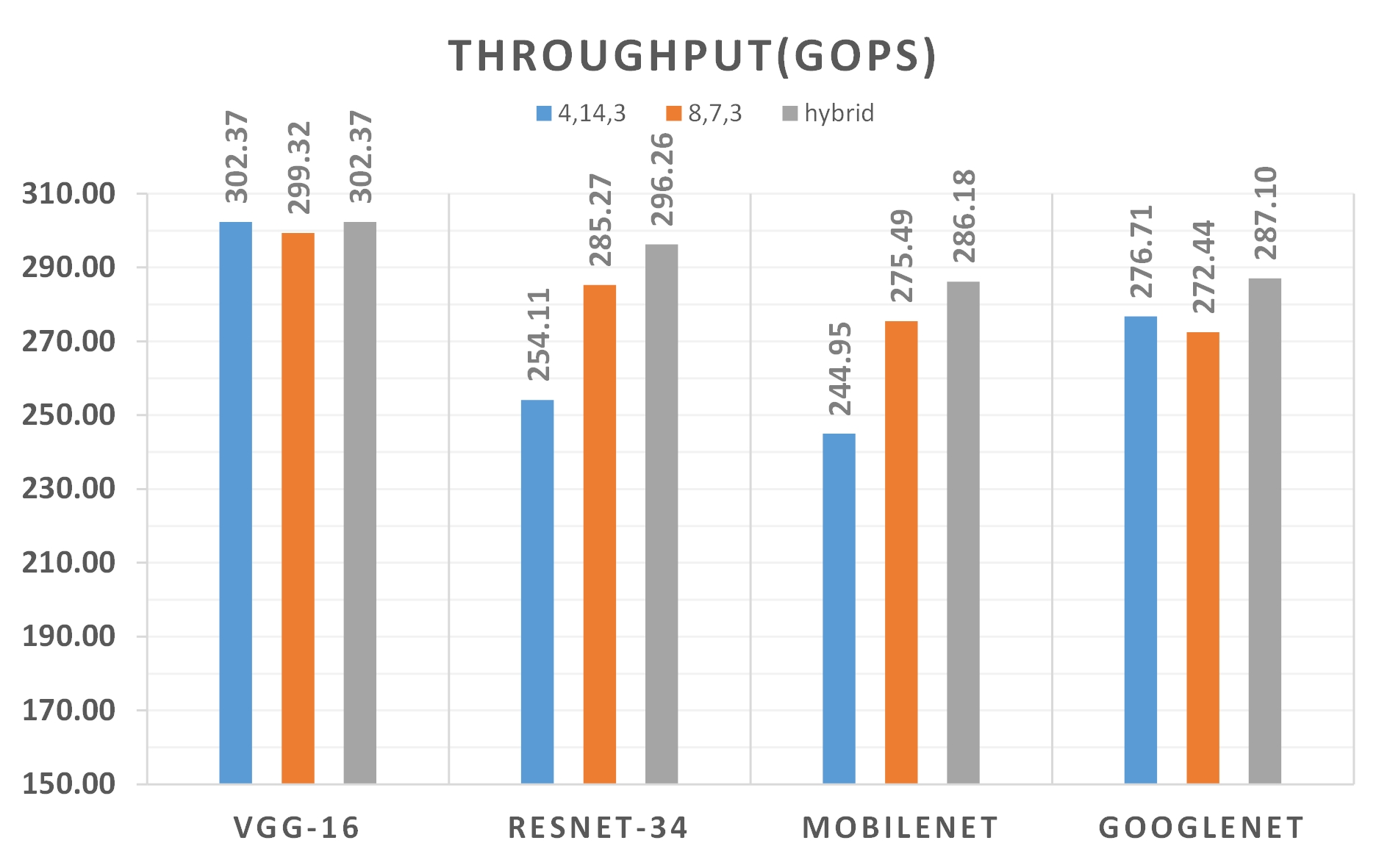}\label{fig:throughput}}
	\caption{Performance in different models and PE configurations.}
	\label{fig:PE_CON}
\end{figure*}	
 \begin{table*}[t]
	\centering
	\caption{Implementation result and comparisons with other designs.}
	\label{table:overview}
	\begin{tabular}{|l||c|c|c|c|c|c|c|c|c|}
		\hline
		& Our work & \cite{eyeriss} & \cite{eyerissv2} &\cite{dna}&  \cite{envision} & \cite{mcc}  &\cite{MAERI} & \textcolor{black}{\cite{R1}}&\textcolor{black}{\cite{R2}} \\\hline
		Technology &40nm & 65nm& 65nm&65nm &28nm  & 65nm &28nm & \textcolor{black}{65nm}&\textcolor{black}{65nm} \\\hline
		\textcolor{black}{Measurements} & \textcolor{black}{Post-layout}&\textcolor{black}{Chip}&\textcolor{black}{Post-layout}&\textcolor{black}{Post-layout}&\textcolor{black}{Chip}&\textcolor{black}{Post-layout}&\textcolor{black}{Post-layout}&\textcolor{black}{Chip}&\textcolor{black}{Post-layout}\\\hline
		\textcolor{black}{Supply Voltage (V)} & \textcolor{black}{0.99} & \textcolor{black}{1.0} & \textcolor{black}{1.0} & \textcolor{black}{1.2} & \textcolor{black}{0.65 - 1.1 }  & \textcolor{black}{1.0}  &\textcolor{black}{-}&\textcolor{black}{1.2}&\textcolor{black}{1.2}\\\hline
		Precision (bits) &16 fixed &16 fixed& 8-20 fixed&16 fixed &4, 8, 16 Dyna.  & 16 fixed &  16 fixed&\textcolor{black}{10}&\textcolor{black}{16 fixed}\\\hline		
		PE number & 168 & 168 &192& 512 &-& 168&168 / 374&\textcolor{black}{128}&\textcolor{black}{64}\\\hline
		Clock rate (MHz)&500&200&200&200&200&500&200&\textcolor{black}{100}&\textcolor{black}{200}\\\hline
		\multirow{2}{3cm}{$^{a}$Peak Throughput (GOPS)}&168&67.2&153.6&204.8&102 - 408&152&67.2 / 149.6 &\textcolor{black}{25.6}&\textcolor{black}{23.4}\\
		&$^{c}$168&$^{c}$109.2&$^{c}$249.6&$^{c}$332.8&$^{c}$71.4 - 285.6&$^{c}$247&$^{c}$47 / 104.7&\textcolor{black}{$^{c}$41.6}&\textcolor{black}{$^{c}$38.0}\\\hline
		$^{b}$Area (KGE) (logic only)&266.9&1176&2695&-&-&1300&- / -&\textcolor{black}{2103}&\textcolor{black}{282}\\\hline
		Area (mm$^{2}$)&$^{e}$1.5625&$^{e}$12.25&-&$^{f}$16.0&$^{e}$1.87&$^{e}$5.0&3.84 / 6.0 &\textcolor{black}{10.52}&\textcolor{black}{10.6}\\\hline
		SRAM(KB)&191&181.5&192&280&144&96&80.5&\textcolor{black}{312}&\textcolor{black}{139.6}\\\hline
		\multirow{2}{3cm}{$^{d}$Area eff. (GOPS/KGE)}&0.629&0.057&0.057&-&-&0.117&- / -&\textcolor{black}{0.012}&\textcolor{black}{0.08}\\
		&\bf{$^{c}$0.629}&$^{c}$0.092&$^{c}$0.092&-&-&$^{c}$0.19&- / -&\textcolor{black}{$^{c}$0.019}&\textcolor{black}{$^{c}$0.13}\\\hline
		\multirow{2}{3cm}{$^{d}$Area eff. (GOPS/mm$^{2}$)}&107.52&5.485&-&12.8&54.5 - 218&30.4&17.5 / 24.9&\textcolor{black}{2.43}&\textcolor{black}{2.20}\\
		&\bf{$^{c}$107.52}&$^{c}$8.914&-&20.8&$^{c}$38.1 - 152&$^{c}$49.4&$^{c}$12.2 / 17.5 &\textcolor{black}{$^{c}$3.95}&\textcolor{black}{$^{c}$3.58}\\\hline	
		$^{g}$Power (mW) &154.98&278&460.5&479&$^{i}$7.5 - $^{j}$300&354&375 / 535&\textcolor{black}{20.5}&\textcolor{black}{93.4}\\\hline
		\multirow{2}{3cm}{Power eff. (TOPS/W)}&1.084&0.241&0.333&0.406&$^{j}$0.26 - $^{i}$10&0.429&0.18 / 0.27&\textcolor{black}{1.24}&\textcolor{black}{0.25}\\
		&\bf{$^{h}$1.084}&$^{h}$0.399&$^{h}$0.552&$^{h}$0.969&$^{h}$$^{j}$0.56 - \bf{$^{i}$3.017}&$^{h}$0.711&-&\textcolor{black}{\bf{$^{h}$2.96}}&\textcolor{black}{$^{h}$0.596}\\\hline
		\multicolumn{4}	{|l}{$^{a}$1 GMACS= 2 GOPS} & \multicolumn{6}	{l|}{$^{b}$The area is shown in terms of the size of kilo NAND2 gates \textcolor{black}{(KGE)}.} \\
		\multicolumn{4}	{|l}{$^{c}$Technology scaling ($\dfrac{process}{40nm}$)} & \multicolumn{6}	{l|}{$^{d}$We take the theoretical performance to evaluate area efficiency fairly here.} \\
		\multicolumn{4}	{|l}{$^{e}$Core only size.} & \multicolumn{6}	{l|}{$^{f}$Chip size.} \\	
		\multicolumn{4}	{|l}{$^{g}$Core only power.} & \multicolumn{6}	{l|}{$^{h}$\textcolor{black}{Normalized~power~efficiency $=$ power~efficiency $\times(\dfrac{process}{40nm})\times(\dfrac{Voltage}{0.99V})^2$.}} \\
		\multicolumn{4}	{|l}{$^{i}$30\% - 60\% sparsity, 3 - 4bits, and 76 GOPS at 0.65V.} & \multicolumn{6}	{l|}{$^{j}$16 bits precsion and 76 GOPS at 1.1V.} \\
		\hline		
	\end{tabular}
\end{table*}
\begin{table*}[t]
	\centering
	\caption{Layer by layer comparison for AlexNet.}
	\label{table:alexnet}
	\begin{tabular}{|l||c|c||c||c||c|}
		\hline
		
		& Our work & \cite{eyeriss} & $^{a}$Ratio &\cite{dna} & $^{a}$Ratio \\\hline
		PE number & 168 & 168 &$^{b}$1&512&$^{b}$0.328 \\\hline
		Clock rate&500MHz&200MHz&$^{b}$2.5&200MHz&$^{b}$2.5\\\hline
		
		Peak throughput~(clock rate * PE) &168&67.2 &$^{b}$2.5& 204.8&$^{b}$ 0.82\\\hline
		Real throughput (GOPS)&146.33&46.314&$^{b}$3.159&180.4&$^{b}$0.811\\\hline

		CONV1~(ms)&1.515&5.225&$^{a}$3.449&1.09&$^{a}$0.719\\\hline
		
		CONV2~(ms)&3.348&10.475&$^{a}$3.129&2.4&$^{a}$0.717\\\hline
		CONV3~(ms)&1.92&5.9&$^{a}$3.073&1.73&$^{a}$0.901\\\hline
		CONV4~(ms)&1.44&4.6&$^{a}$3.194&1.3&$^{a}$0.903\\\hline
		CONV5~(ms)&0.96&2.625&$^{a}$2.734&0.86&$^{a}$0.896\\\hline
		Total Latency~(ms)&9.184&28.825&$^{a}$3.139&7.38&$^{a}$0.804\\\hline
		\multicolumn{6}	{|l|}{$^{a}$Ratio = compablack work / our work for layer latency.}\\ 
		\multicolumn{6}	{|l|}{$^{b}$Ratio = our work / compablack work for other.}\\\hline
	\end{tabular}
\end{table*}
\begin{table}[t]
	\centering
	\caption{Layer by layer comparison for VGG-16.}
	\label{table:VGG}
	\begin{tabular}{|c||c|c||c|}
		\hline
		
		& Our work & \cite{eyeriss} & Ratio \\\hline
		PE number & 168 & 168&$^{b}$1 \\\hline
		Clock rate&500MHz&200MHz&$^{b}$2.5\\\hline
		Peak throughput (clock rate * PE) & 168&67.2 & $^{b}$2.5\\\hline
		CONV1 (ms)&5.161&25.4&$^{a}$4.922\\\hline
		CONV2 (ms)&11.012&303.43&$^{a}$27.555\\\hline
		CONV3 (ms)&5.506&156.77&$^{a}$28.473\\\hline
		CONV4 (ms)&11.012&298.1&$^{a}$27.071\\\hline
		CONV5 (ms)&5.506&80.37&$^{a}$14.597\\\hline
		CONV6 (ms)&11.012&153.63&$^{a}$13.951\\\hline
		CONV7 (ms)&11.012&152.57&$^{a}$13.855\\\hline
		CONV8 (ms)&5.506&45.27&$^{a}$8.222\\\hline
		CONV9 (ms)&11.012&84.93&$^{a}$7.713\\\hline
		CONV10 (ms)&11.012&82.1&$^{a}$7.456\\\hline
		CONV11 (ms)&2.75&18.1&$^{a}$6.509\\\hline
		CONV12 (ms)&2.75&17.9&$^{a}$6.509\\\hline
		CONV13 (ms)&2.75&17.9&$^{a}$6.509\\\hline
		Total Latency (ms)&100.69&1436.47&$^{a}$14.266\\\hline
		\multicolumn{4}	{|l|}{$^{a}$Ratio = compablack work / our work for layer latency.}\\ 
		\multicolumn{4}	{|l|}{$^{b}$Ratio = our work / compablack work for throughput.}\\\hline
		
	\end{tabular}
\end{table}
\subsection{Area Analysis}
Fig.~\ref{fig.area}(a) shows area breakdown of the core. Without SRAM, PE array occupies 70\% of \textcolor{black}{total} area due to 168 MACs while accumulator occupies 29\%. Area of the controller is the lowest due to the simple and regular data flow. For the whole core area in Fig.~\ref{fig.area}(b), SRAM buffers occupies 78.6\% due to large SRAM buffer for data reuse. 

\subsection{\textcolor{black}{Analysis of Hardware Utilization and Performance}}
Fig.~\ref{fig:vgg_u} to \ref{fig:google} show hardware utilization of each layer in VGG-16, Resnet-34, MobileNet, and GoogLeNet with two PE configurations: (8, 7, 3) or (4, 14, 3). In which, (3$\times$3/s1 3/64) as shown in Fig.~\ref{fig:vgg_u} to \ref{fig:google} represents 3$\times$3 convolutions with unit stride, 3 input channels, and 64 output channels. 

For VGG-16, it consists of all 3$\times$3 convolutions and thus has 100\% utilization for all layers except the first layer. The first layer only has 75\% utilization for (4, 14, 3) since only 3 instead of 4 blocks are used to compute 3 channel input. 

For ResNet-34, the first layer is 7$\times$7 convolutions with stride 4, which can be decomposed into 4$\times$4 convolutions with unit stride. However, 4$\times$4 convolutions do not achieve 100\% of hardware utilization. Thus, the first layer will have only 77\% of utilization for the (4, 14, 3) case or 38\% of utilization for the (8, 7, 3) case. Most of the middle layers achieve 100\% of utilization for its 3$\times$3 convolutions except for the layers of the 3$\times$3 convolution with stride 2 that have only 80\% of utilization. The  \textcolor{black}{utilization of the} last layer of the 3$\times$3 convolutions with stride 2 is dropped to 58\% due to small 7$\times$7 feature maps. The final few layers have 49\% of utilization for the (4. 14, 3) case due to small feature map size, 7, in large PE rows, 14, of a block. To improve \textcolor{black}{hardware} utilization, we reconfigure the PE array to be (4, 14, 3) for the first layer, and (8, 7, 3) for the last few layers with the layer adaptive PE configuration, and thus have the best one from (8, 7, 3) and (4, 14, 3) cases. 

For MobileNet, the layers with 3$\times$3 stride 1 have 100\% of utilization. 1$\times$1 layers have 89\% of utilization since we only process $ 8\times3=24 $ or $ 4\times3= 12$ channels at a time and cannot have full hardware utilization for the certain number of channels. The layer with 3$\times$3 stride 2 has lower hardware utilization with the same reason as above. In this case, we can also use layer adaptive PE reconfiguration as above for higher utilization. In summary, with the proposed data flow, we can attain high utilization while keep structure simple.

For GoogLeNet, utilization of the most layers with 3$\times$3 and 1$\times$1 achieves near 100\% of utilization due to our reconfigurable PE array for different convolution kernels. Only the first one and last few layers have lower utilization due to the same reasons as in MobileNet. However, with our configurable PE blocks, we can attain higher utilization for these layers.

Fig.~\ref{fig:utilization} shows the overall hardware utilization in different models and PE configurations. The utilization of 4 blocks is always lower than 8 blocks because hardware utilization is only half when the size of the feature map is down-sampled to 7 (half of 14 rows) for ResNet-34, GoogLeNet, and MobileNet.  VGG-16 is the only exception since its minimum feature map size is 14. The layer adaptive PE configuration has the best utilization. With the proposed data flow, the utilization can achieve 99\%, 97\%, 93.7\%, and 94\% for VGG-16, ResNet-34, MobileNet and GoogLeNet, respectively.

Fig.~\ref{fig:throughput} shows the overall throughput in different models and PE configurations. These throughput numbers are proportional to the hardware utilization since our models have optimized for different convolutional types in these models.

\subsection{Design Comparison}
Table.~\ref{table:overview} shows the implementation result and comparison with other designs. The proposed design has much lower area cost than other designs due to the simpler PE structure and regular local summation. The peak throughput can reach up to 168 GOPS. The area efficiency is 0.629 GOPS/KGE and 107.52 GOPS/mm$^{2}$, \textcolor{black}{which} is much higher than other designs. The power consumption is 154.98mW and power efficiency can reach up to 1.084 TOPS/W because of the simple and regular data flow. The power efficiency is better than most of designs except for \cite{envision} and \cite{R1}. \textcolor{black}{\cite{envision} uses lower bitwidth hardware to compute 30\% - 60\% sparse network at lower supply voltage to attain lower area cost or power consumption. \cite{R1} uses an optimized numerical representation for lower bitwidth hardware to achieve lower power consumption.}

Table ~\ref{table:alexnet} shows a layer-by-layer performance comparison for AlexNet. For a fair comparison, we first list speedup ratio of the peak throughput between different designs and the proposed work. \textcolor{black}{This can} normalize the effects of different clock rate and PE numbers \textcolor{black}{in each design}, which implies a full hardware utilization case. Then, for each layer, if the compablack design has lower hardware utilization than the proposed one, the speedup ratio will become larger. Otherwise, the speedup ratio will be lower. By dividing speedup of the layers with speedup of the peak throughput, we will know the difference in hardware utilization between designs. Thus, compablack to \cite{eyeriss}, the hardware utilization of our design is 20.4\% $(3.139-2.5/3.139)$ higher. When compablack to \cite{dna}, the hardware utilization of our design 1.95\% $(0.804-0.82/0.82)$ lower since our design has lower hardware utilization for CNN with low number of channels and non-3$\times$3 convolutions at the first and second layers of AlexNet, respectively. However, this difference is quite small. With our 3$\times$3 optimized based structure, we can have the simpler structure and lower cost compablack to \cite{dna}. Beyond the first two layers, \textcolor{black}{the utilization of} our design is 12\% higher at other layers, which proves the efficiency of the proposed design.

Table ~\ref{table:VGG} shows another comparison for VGG-16. The \textcolor{black}{the utilization of} proposed design has 470\% higher hardware utilization than \textcolor{black}{that in} \cite{eyeriss}. The reason is that these layers are our optimized 3$\times$3 layers \textcolor{black}{and have} 100\% of hardware utilization. The only exception is the first layer due to only three input channels. The large hardware utilization degradation at CONV1 to CONV7 in \cite{eyeriss} is due to bad handling of the large feature map in their PE structure. On the contrary, the broadcasted vectorwise input approach in our design can keep PE hardware utilization high.

\section{Conclusion}
This paper proposes a hardware efficient CNN accelerator that can execute different convolution filters efficiently, such as 3$\times$3, 4$\times$4, 5$\times$5, 3$\times$3 stride 2, depthwise, and 1$\times$1 convolutions. This flexible scheduling is enabled by layer adaptive PE configurations, interleaved input selection scheme, elementwise input, and reconfigurable accumulators. To blackuce cost but still meet above flexibility requirements, we adopt a systolic array architecture with vectorwise input and weight data to keep structure regular. The proposed design can achieve at least 93.7\% of hardware utilization \textcolor{black}{for commonly used image recognition networks}, with 266.9K NAND gate counts, 191KB SRAM buffer, and 1.25mm $\times$ 1.25mm core area when implemented on TSMC 40nm CMOS process, which has higher efficiency than existing designs.

\section*{Acknowledgment}
This work was supported by Ministry of Science and Technology, Taiwan, under Grant 108-2634-F-009-005, 107-2119-M-009-019, and Research of Excellence program 106-2633-E-009-001.

\vspace{50 mm} 
\begin{IEEEbiography}
	[{\includegraphics[width=1in,height=1.25in,clip,keepaspectratio]{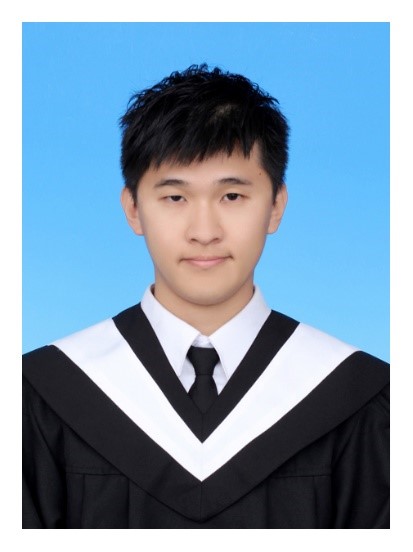}}]{Kuo-Wei, Chang}
	is currently pursuing the Ph.D. degree with the National Chiao Tung University(NCTU), Hsinchu, Taiwan, R.O.C., and focus on VLSI design and hardware implementation for deep learning accelerators.
\end{IEEEbiography}
\vspace{160 mm}
\begin{IEEEbiography}
	[{\includegraphics[width=1in,height=1.25in,clip,keepaspectratio]{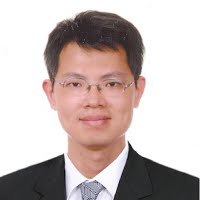}}]
	{Tian-Sheuan Chang}
	(S’93–M’06–SM’07)
	received the B.S., M.S., and Ph.D. degrees in electronic
	engineering from National Chiao-Tung University (NCTU),
	Hsinchu, Taiwan, in 1993, 1995, and 1999, respectively.
	From 2000 to 2004, he was a Deputy Manager with Global
	Unichip Corporation, Hsinchu, Taiwan. In 2004, he joined the
	Department of Electronics Engineering, NCTU, where he is
	currently a Professor. In 2009, he was a visiting scholar in
	IMEC, Belgium. His current research interests include systemon-
	a-chip design, VLSI signal processing, and computer
	architecture.
	Dr. Chang has received the Excellent Young Electrical
	Engineer from Chinese Institute of Electrical Engineering in
	2007, and the Outstanding Young Scholar from Taiwan IC
	Design Society in 2010. He has been actively involved in many
	international conferences as an organizing committee or
	technical program committee member.
\end{IEEEbiography}

\end{document}